\documentclass{article} 

\usepackage{olo}
\usepackage{latexsym}
\usepackage{fullpage}
\usepackage{natbib}

\newcommand{\mymodel}{\textsf{VIPER}\xspace}
\newcommand{\myalgo}{\textsf{ESOP}\xspace}
\newcommand{\epi}{\text{epi}}
\newcommand{\eco}{\text{eco}}
\newcommand{\vxo}{\vx^\ast}

\graphicspath{{figs/}{figs/case/}}

\begin{document}

\title{Epidemiologically and Socio-economically Optimal Policies via Bayesian Optimization}
\author{Amit Chandak \and Debojyoti Dey \and Bhaskar Mukhoty \and Purushottam Kar\\Indian Institute of Technology Kanpur\\\texttt{\{amitch,debojyot,bhaskarm,purushot\}@cse.iitk.ac.in}}
\date{\today}

\maketitle

\begin{abstract}
Mass public quarantining, colloquially known as a \emph{lock-down}, is a non-pharmaceutical intervention to check spread of disease. This paper presents \myalgo (Epidemiologically and Socio-economically Optimal Policies)\footnote{All code used for this study is available at the following GitHub Repository\newline\url{https://github.com/purushottamkar/esop}}, a novel application of active machine learning techniques using Bayesian optimization, that interacts with an epidemiological model to arrive at lock-down schedules that optimally balance public health benefits and socio-economic downsides of reduced economic activity during lock-down periods. The utility of \myalgo is demonstrated using case studies with \mymodel (Virus-Individual-Policy-EnviRonment), a stochastic agent-based simulator that this paper also proposes. However, \myalgo is flexible enough to interact with arbitrary epidemiological simulators in a black-box manner, and produce schedules that involve multiple phases of lock-downs.\\
\emph{\textbf{Disclaimer}: This paper makes no recommendation to individuals and its results should not be interpreted by individuals to modulate personal behavior. The authors recommend that individuals continue to follow guidelines offered by local governments with respect to lock-downs and social distancing, and those offered by medical professionals with respect to personal hygiene and treatment.
}
\end{abstract}

\section{Introduction}
\label{sec:intro}
Infectious diseases that are contagious pose a threat to public safety once they attain pandemic status. Several historical instances of such pandemics have taken a heavy toll on human lives. Prominent examples include the H1N1 (Spanish flu) pandemic of 1918 ($>$ 50 million fatalities), the H3N2 (HongKong flu) pandemic of 1968 ($\approx$ 1 million fatalities), the HIV/AIDS pandemic ($\approx$ 32 million fatalities) \citep{KimballBose2020}, the novel influenza-A H1N1 (swine flu) pandemic of 2009 ($\approx$ 0.3 million fatalities) \citep{Roos2012}, and the ongoing CoViD-19 pandemic ($\approx$ 0.4 million fatalities as of writing this document) \citep{WHO2020}.

In such situations, and especially in the absence of vaccines and antiviral treatments, experts often prescribe guidelines to public such as hand hygiene and respiratory etiquette, as well as two kinds of non-pharmaceutical interventions, namely 1) \emph{mitigation policies} such as human surveillance and contact tracing, and 2) \emph{suppression policies} such as social distancing or its more extreme form colloquially known as a \emph{lock-down} \citep{AledortLWB2007}. However, their benefits with respect to public health outcomes notwithstanding, severe and extended applications of suppression policies such as lock-downs negatively impact livelihoods and the economy. For instance, \cite{Scherbina2020} estimates the cost of extensive suppression measures to the US economy at \$9 trillion, or about 43\% of its annual GDP.

\cite{PeakCGB2017} demonstrate the need for policy decisions to balance suppression and mitigation measures in terms of the epidemiological characteristics of the pandemic, pointing out that suppression measures hold most benefit for fast-course diseases whereas effective mitigation measures may suffice for others at much less socio-economic cost. This points to a need for techniques that can take the disease progression characteristics of a certain outbreak and suggest policies that optimally use suppression and mitigation techniques to offer acceptable health outcomes as well as socio-economic risks within acceptable limits. In fact there is an emerging area variously termed ``economic epidemiology'' or ``epidemiological economics'' \cite{Perringsetal2014} that seeks to develop models that can address the interplay of disease and host behavior. 

\subsection{Related Works}
\label{sec:related}
Several works exist on modelling epidemic and pandemic progressions using differential equation-based models such as SIR or SEIR and using them to make predictions. Some examples include \citep{EfimovUshirobira2020,LyraNJBACdA2020,SardarNC2020,VyasarayaniChatterjee2020}. Most of these studies utilize differential equation-based models such as SIR and SEIR variants. This paper instead uses a stochastic agent-based model called \mymodel that is proposed in Sec~\ref{sec:problem}. The reason behind this choice was to demonstrate the effectiveness of our proposed techniques when working with epidemiological models that are not described compactly using a few equations and thus, harder to analyze.

Micro-simulation studies using UK \citep{Fergusonetal2020} and Indian \citep{SinghAdhikari2020} data conclude that multiple short-term suppression rounds may offer acceptable health outcomes when a single extended period of suppression is infeasible. However, these works do not offer ways to find either the optimal moment to initiate suppression measures or their duration.

This is important since \cite{MorrisRPL2020,Patterson-Lomba2020} show that the optimal initiation point and duration of a suppression may depend substantially on the disease and social characteristics themselves (for instance the disease incubation period and the basic reproduction number $R_0$). This is understandable since premature suppression would slow down the depletion of the pool of susceptible individuals leaving room open for a second wave of infections whereas delayed suppression may cause the initial wave to be widespread in itself. \cite{Scherbina2020} additionally considers the economic impact of these measures and suggests durations for lock-down periods and their associated economic costs in medical expenses as well as lost value of statistical life.

Prior works offering actual policy advice fall into two categories: 1) those that offer only broad principles on how to target interventions e.g. by identifying simple rules of thumb \citep{WallingaBL2010}, and 2) those that do offer actionable advice e.g. when to initiate suppression \citep{KlepacLG2011,MorrisRPL2020,TorreMM2019,ZhaoFeng2019}. However, the latter often do not take the socio-economic impact of these measures into account and moreover, consider only simple theoretical models e.g. SIR that are not very expressive.

A subclass of the latter approaches \citep[for example,][]{BussellDGC2019} advocate first fitting an approximate model to the actual simulator (to make it simple enough to enable mathematical analyses) before applying optimal control strategies. Such approximations may introduce unmodelled errors into the prediction pipeline and adversely affect their outcome. \myalgo instead directly models intervention outcomes in terms of the simulator outputs.

\subsection{Our Contributions}
\label{sec:contrib}
This paper presents \myalgo (Epidemiologically and Socio-economically Optimal Policies), a system that uses Bayesian optimization to automatically suggest suppression policies that optimally balance public health and economic outcomes. \myalgo interacts with epidemiological and economic models to automatically suggest policy decisions. The paper also presents \mymodel (Virus-Individual-Policy-EnviRonment), an iterative, stochastic agent-based model (ABM) with which case studies are conducted to showcase the utility of \myalgo. We note however, that \myalgo can readily interact with other epidemiological models, e.g. those that incorporate stratification based on region and age e.g. INDSCI-SIM \citep{Shekatkaretal2020}, COVision \citep{Nagorietal2020}, ABCS \citep{Harshaetal2020} and IndiaSim \citep{Megiddoetal2014}. Although machine learning techniques have been used in epidemiological forecasting \citep{LindstromTW2015} and estimating model parameters \citep{DandekarBarbastathis2020}, we are not aware of prior work using machine learning in epidemiological policy design.

\begin{table}[t]%
\setlength{\tabcolsep}{1ex}
\resizebox{0.5\textwidth}{!}{%
\begin{tabular}{clcc}
Attr & Description                       & Range  & Def    \\\hline
\multicolumn{3}{c}{\textbf{Viral Model}}                       \\
\hline
INC      & incubation period                & $\bN$ & 3       \\
BVL      & base viral load                  & $[0,1]$ & 0.05     \\
DPR      & disease progression rate         & $[0,1]$ & 0.1     \\
XTH      & VLD threshold for expiry & $[0,1]$ & 0.7     \\
BXP      & expiry probability at XTH       & $[0,1]$ & 0.0     \\\hline
\multicolumn{3}{c}{\textbf{Environment Model}}                       \\
\hline
BCR      & contact radius b/w individuals      & $[0,1]$ & 0.25     \\
BIP      & prob. infection upon contact   & $[0,1]$ & 0.5     \\
BTR      & prob. of an individual traveling         & $[0,1]$ & 0.01     \\
BTD      & maximum travel distance                     & $[0,1]$ & 1.0     \\
INI      & initial rate of infection    & $[0,1]$ & 0.01     \\\hline
\end{tabular}
}
\resizebox{0.5\textwidth}{!}{%
\begin{tabular}{clcc}
Attr & Description                       & Range  & Ini     \\\hline
\multicolumn{3}{c}{\textbf{Individual Model}}                           \\
\hline
SUS      & susceptibility to infection       & $[0,1]$ & rnd         \\
RST      & resistance to disease progression    & $[0,1]$ & rnd          \\
VLD      & current viral load                & $[0,1]$ & 0.0         \\
RLD      & current recovery load               & $[0,1]$ & 0.0          \\
STA      & current state                     & SEIRX & S \\
QRN      & quarantine status                 & $0$ or $1$ & 0       \\
X, Y     & current location                  & $[0,1]^2$ & rnd          \\\hline
\multicolumn{3}{c}{\textbf{Policy Model}}                       \\
\hline
QTH      & VLD threshold for quarantine & $[0,1]$ & 0.3     \\
BQP      & quarantine probability at QTH       & $[0,1]$ & 0.0     \\
$l(t)$   & lock-down level at time $t$        & $[0,5]$ & ---		\\\hline
\end{tabular}
}
\caption{\mymodel model attributes, valid ranges and default/initial values. See Sec~\ref{sec:problem} for details.}
\label{tab:models}
\end{table}

\section{\mymodel: An Iterative Stochastic Agent-based Epidemiological Model}
\label{sec:problem}
\mymodel (Virus-Individual-Policy-EnviRonment) models an \emph{in-silico} population of individuals, supports compartments of the SEIR model \citep{KeelingRohani2008}, and allows travel and quarantining of individuals. Being an ABM rather than an ODE-based model, \mymodel can model disease progression within each individual separately and thus, quarantine or expire individuals based on their stage of the disease, something that is difficult to do in ODE-based models. Stochastic ABMs allow diverse socio-medico-economic traits to be modeled at the individual level but cannot be easily represented by a concise system of ODEs. Thus, works such as \citep{MorrisRPL2020} do not apply here.

Details of the \mymodel model are described below and succinctly enumerated in Tab~\ref{tab:models}. \mymodel consists four \emph{sub-models}, one each devoted to modelling individuals, the virus, environment parameters and policy parameters.

\paragraph{Individual Model}: an individual is characterized by their susceptibility to infection (SUS), resistance to disease progression (RST), viral load (VLD), recovery load (RLD), current state (STA), quarantine status (QRN) and location (X,Y). SUS, RST, VLD DLD, X and Y are real numbers between $0$ and $1$, whereas QRN takes Boolean values. The state of an individual STA can be either S (susceptible), E (exposed), I (infectious), R (recovered) or X (expired/deceased). Individuals are initialized with random values for RST, SUS and their location within the 2-D box $[0,1]^2$. Individuals progress from state S $\rightarrow$ E $\rightarrow$ I, can be optionally quarantined while in state I, and then move to either state R or X.

\paragraph{Viral Model}: the virus is characterized by its incubation period (INC), the base viral load in an individual at the end of the incubation period (BVL), the disease progression rate (DPR), the viral load over which an individual's chances of getting expired start increasing (XTH) and the base removal probability of an individual with viral load at XTH (BXP). BVL, DPR, XTH and BXP are real numbers between $0$ and $1$ whereas INC is a natural number.

\paragraph{Environment Model}: the environmental factors are modeled using the typical contact radius between individuals (BCR), the probability that a contact between an infectious and susceptible individual will lead to a successful infection (BIP), the fraction of the population that travels at any time instant (BTR), the maximum distance to which they travel (BTD), and the fraction of population that is infected with the virus at start of the simulation (INI). All these values are represented as real numbers between $0$ and $1$.

\paragraph{Policy Model}: the policy model comprises the viral load over which an individual's chances of getting quarantined start increasing (QTH) and the base quarantining probability of an individual with viral load at QTH (BQP). Both are real numbers between $0$ and $1$. Additionally, the policy prescribes a \emph{lock-down} level which is a real number between $0$ and $5$. The lock-down level is specified at every time instant $t$ of the simulation. A lock-down level of $l$ at a certain time instant causes BTD as well as BCR to go down by a factor of $\exp(-l)$. Thus, at a high lock-down level, individuals are neither able to travel much, nor interact with other individuals far off from their current location. 

\paragraph{Modelling disease-progression dynamics in \mymodel}: The viral load (VLD) of an individual represents the extent of infection within their system. At the end of the incubation period (INC), an exposed individual always has a ``base'' viral load of BVL. The virus attempts to increase this viral load according to the disease progression rate (DPR) whereas the individual resists this according to their resistance level (RST) by converting viral load to recovery load (RLD). Disease progression in every infected individual is governed by an SIR-like model with
\[
\frac{d\text{ VLD}(t)}{dt} = -\text{RST} \cdot \text{VLD}(t) + \text{DPR} \cdot (1 - \text{VLD}(t) - \text{RLD}(t))
\]
and
\[
\frac{d\text{ RLD}(t)}{dt} = \text{RST} \cdot \text{VLD}(t)
\]
Thus, \mymodel allows individuals to experience disease progression, as well as associated effects like quarantining or expiry, in a completely individualized manner, something that is readily possible in agent-based models but much more difficult to express in terms of a compact set of differential equations.

An infected individual whose VLD falls below BVL moves on to state R, i.e. recovers. An individual with VLD equal to QTH (resp. XTH) has a probability BQP (resp. BXP) of getting quarantined (resp. expired). An individual with VLD greater than QTH has the following probability of getting quarantined
\[
\P{\text{quarantine}} = \text{BQP} + (1 - \text{BQP}) \cdot \frac{\text{VLD} - \text{QTH}}{1 - \text{QTH}},
\]
i.e. the probability of getting quarantined increases linearly to $1$ as the individual's VLD goes up. At every time step $t$, a coin is tossed for all individuals with VLD greater than QTH which lands heads with this particular probability. If the coin does indeed land heads, the individual is deemed quarantined. Note that this allows \mymodel to model asymptomatic transmission since it allows individuals with low VLD levels (esp. below QTH) to avoid detection with high probability but makes it difficult for those in advanced stages of the disease to avoid quarantine.

Similarly, an individual with VLD greater than XTH has the following probability of getting expired
\[
\P{\text{expiry}} = \text{BXP} + (1 - \text{BXP}) \cdot \frac{\text{VLD} - \text{XTH}}{1 - \text{XTH}}.
\]
At every time step, a coin is similarly tossed to decide on whether an individual with VLD greater than XTH gets expired or not. The lock-down level needs to be specified at every time instant $t$ of the simulation. A lock-down level of $l(t)$ causes BTD as well as BCR at that time $t$ to go down by a factor of $\exp(-l(t))$. Thus, at high lock-down levels, individuals are neither able to travel much, nor have contact with other individuals far off from their current location. The lock-down level has no effect on the quarantining or expiry processes described above which continue the same way irrespective of the lock-down level.

\begin{figure}[t]%
\includegraphics[width=0.35\columnwidth]{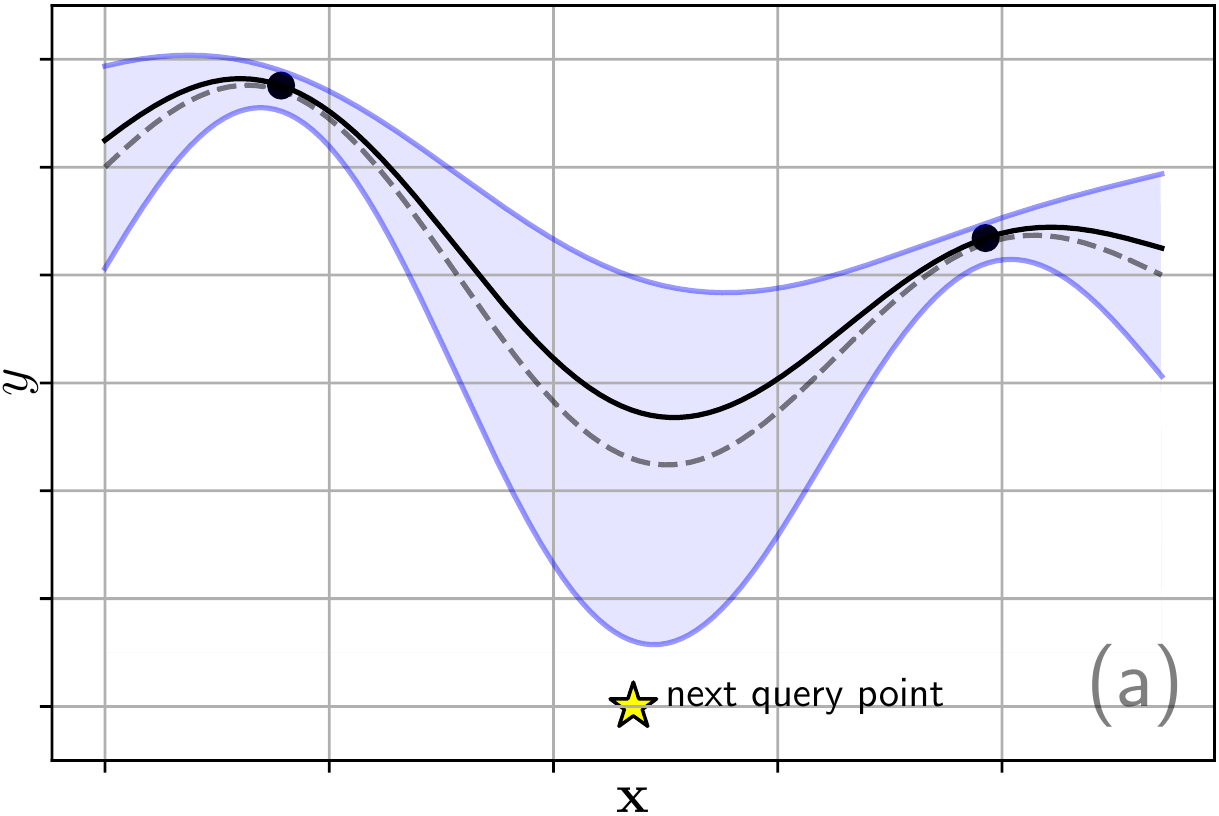}\hfill%
\includegraphics[width=0.35\columnwidth]{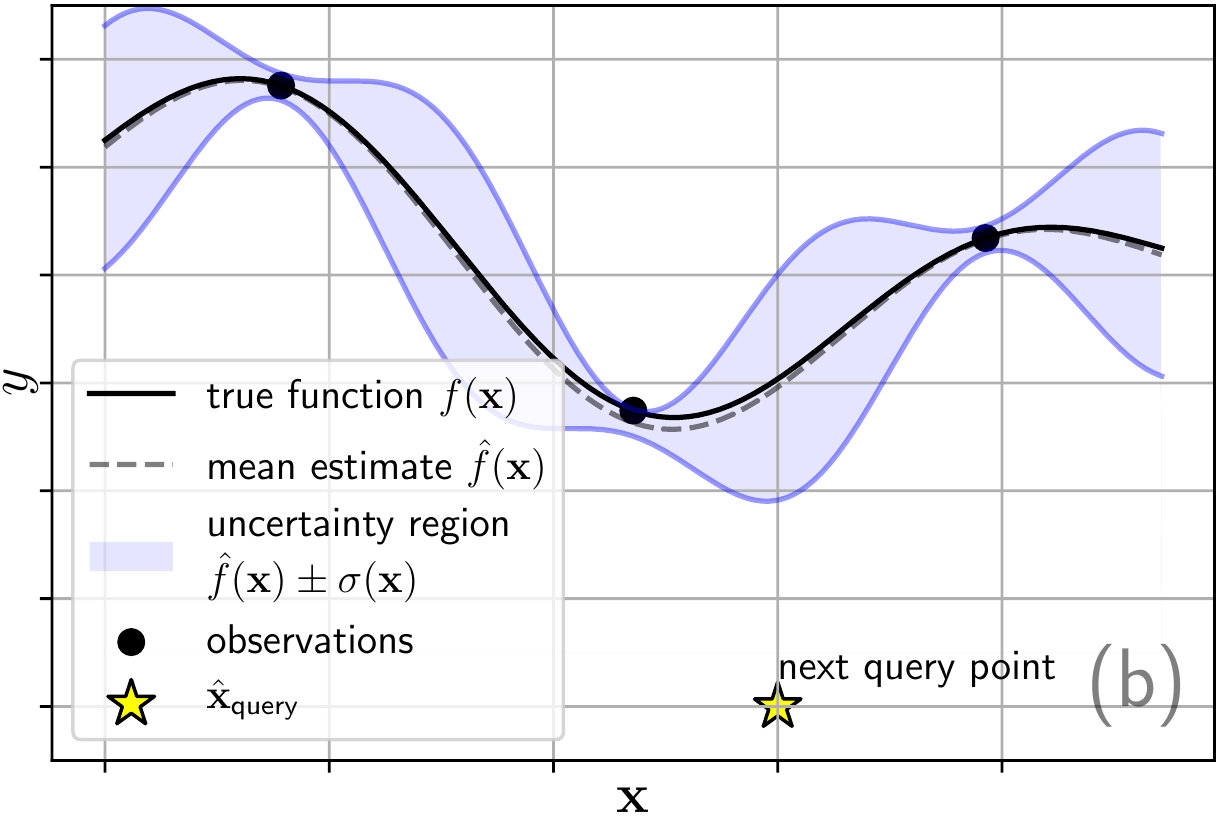}\hfill%
\includegraphics[width=0.26\columnwidth]{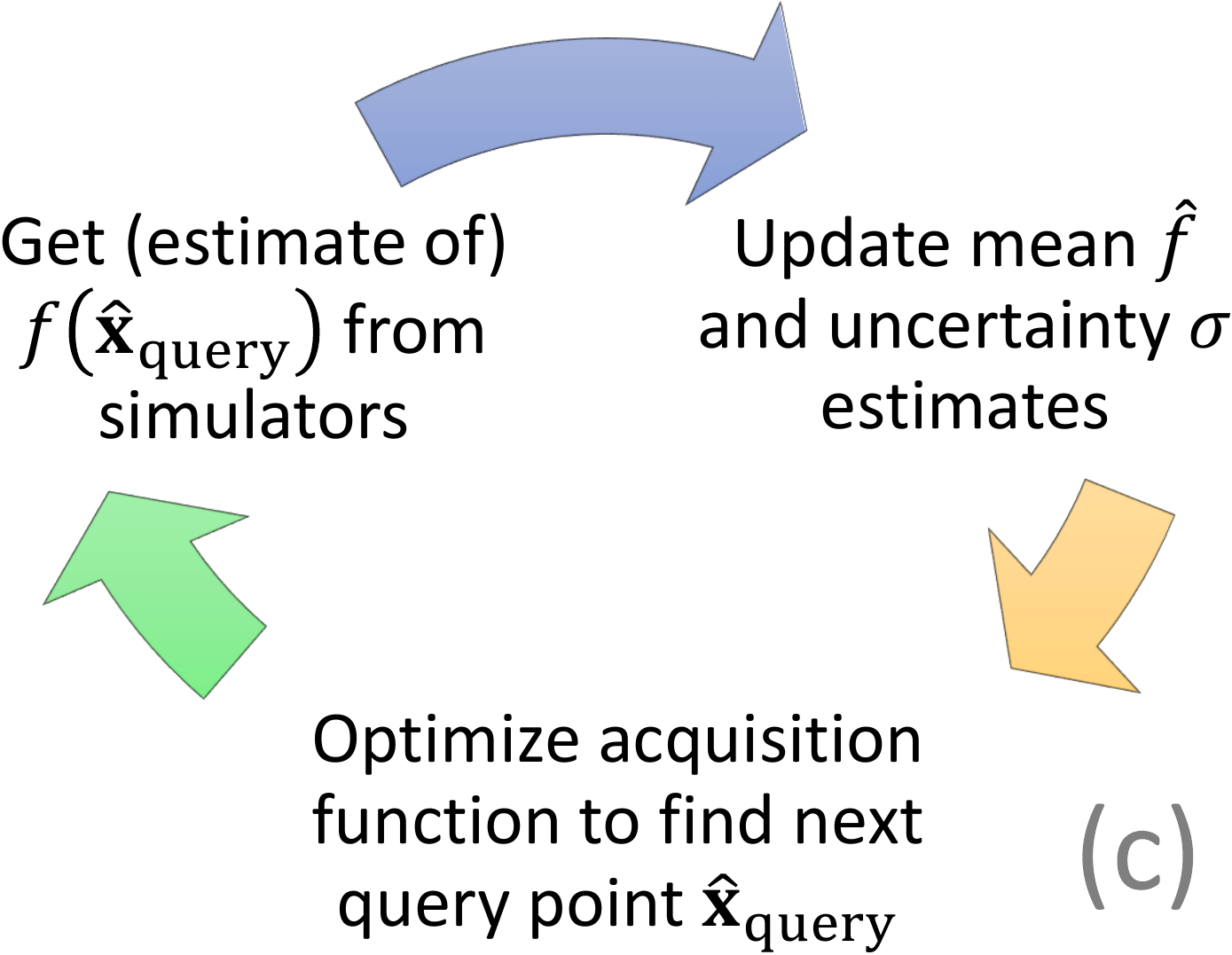}%
\caption{\textbf{An illustration of the Bayesian optimization process}. See Sec~\ref{sec:bo} for details.}%
\label{fig:bo}%
\end{figure}

\section{\myalgo: Epidemiologically and Socio-economically Optimal Policies}
\label{sec:bo}
\myalgo encodes interventions as vectors and their health and socio-economic outcomes as functions, e.g., the coordinates of a 2-D vector $\vx = [\vx_1, \vx_2] \in \bN^2$ may encode the starting point ($\vx_1$) and duration ($\vx_2$) of a lock-down. Next, consider a function $f_\epi: \bN^2 \rightarrow [0,1]$ encoding health outcomes with $f_\epi(\vx)$ equal to the peak infection rate (the largest fraction of the total population infected at any point of time) if the intervention $\vx$ is applied. Similarly, let $f_\eco: \bN^2 \rightarrow [0,1]$ encode economic outcomes with $f_\eco(\vx)$ being the fraction of population that would face unemployment if lock-down were indeed to last $\vx_2$ days. We stress that the functions $f_\epi, f_\eco$ described here are examples and other measurable outcomes, e.g. cumulative death rate, loss to GDP, can also be used. Predicted estimates for $f_\epi$ would be obtained from epidemiological simulators such as INDSCI-SIM or IndiaSim (we will use \mymodel) and those for $f_\eco$ would be obtained from economic models. Our goal is to balance health and economic outcomes by solving the following optimization problem:
\[
\vxo := \arg\min_{\vx \in \bN^2}\ f(\vx) \text{ where } f(\vx) = f_\epi(\vx) + f_\eco(\vx)
\]
However, it is challenging to perform this optimization using standard descent techniques \citep{BoydVandenberghe2003} since even obtaining values of the function $f$ at specific query points (let alone gradients) is expensive as it involves querying simulators such as \mymodel. An acceptable solution in this case is Bayesian optimization \citep{JonesSW1998} which is an \emph{active machine learning} technique used to optimize functions which are expensive to evaluate and to which, moreover we do not have access to gradients.

Fig~\ref{fig:bo} presents a visual depiction of the key processes in Bayesian optimization. The technique adaptively queries the function at only a few locations to quickly approximate the solution to the optimization problem. The algorithm uses the current observations and Gaussian process regression \citep{RasmussenWilliams2006} to obtain a \emph{mean} estimate $\hat f(\vx)$ (dashed line in Fig~\ref{fig:bo}) of the true function $f(\vx)$ (bold line in Fig~\ref{fig:bo}) as well as an estimate $\sigma(\vx)$ of the \emph{uncertainty} in that estimate (the blue shaded region depicts $\hat f(\vx) \pm \sigma(\vx)$ in Fig~\ref{fig:bo}). Notice that uncertainty drops around observation points since (a good estimate of) the true function value is known there.

Using these, an \emph{acquisition} function is created. Fig~\ref{fig:bo}(a) uses the simple LCB (lower confidence bound) acquisition function defined as $a(\vx) = \hat f(\vx) - \sigma(\vx)$. Other possibilities include EI (expected improvement) and KG (knowledge gradient). The (estimated) function value at the point $\hat\vx_\text{query} := \arg\min a(\vx)$ is now queried. Using $f(\hat\vx_\text{query})$, the mean and uncertainty estimates $\hat f,\sigma$ are updated as shown in Fig~\ref{fig:bo}(b) and the process is repeated. At the end, the query point with the lowest function value is returned as the estimated minimum. Note that \myalgo can only query \mymodel but does not have access to its internal attribute values.

Lack of space does not permit a detailed overview and we refer the reader to \citep{Frazier2018} for an excellent review. \myalgo also employs multi-scale search and caching techniques to accelerate computations. Bayesian optimization routines often enjoy provable convergence bounds which we briefly discuss in Sec~\ref{sec:conv-rate}.

\begin{figure}%
\hspace*{0.06\textwidth}\includegraphics[width=0.41\columnwidth]{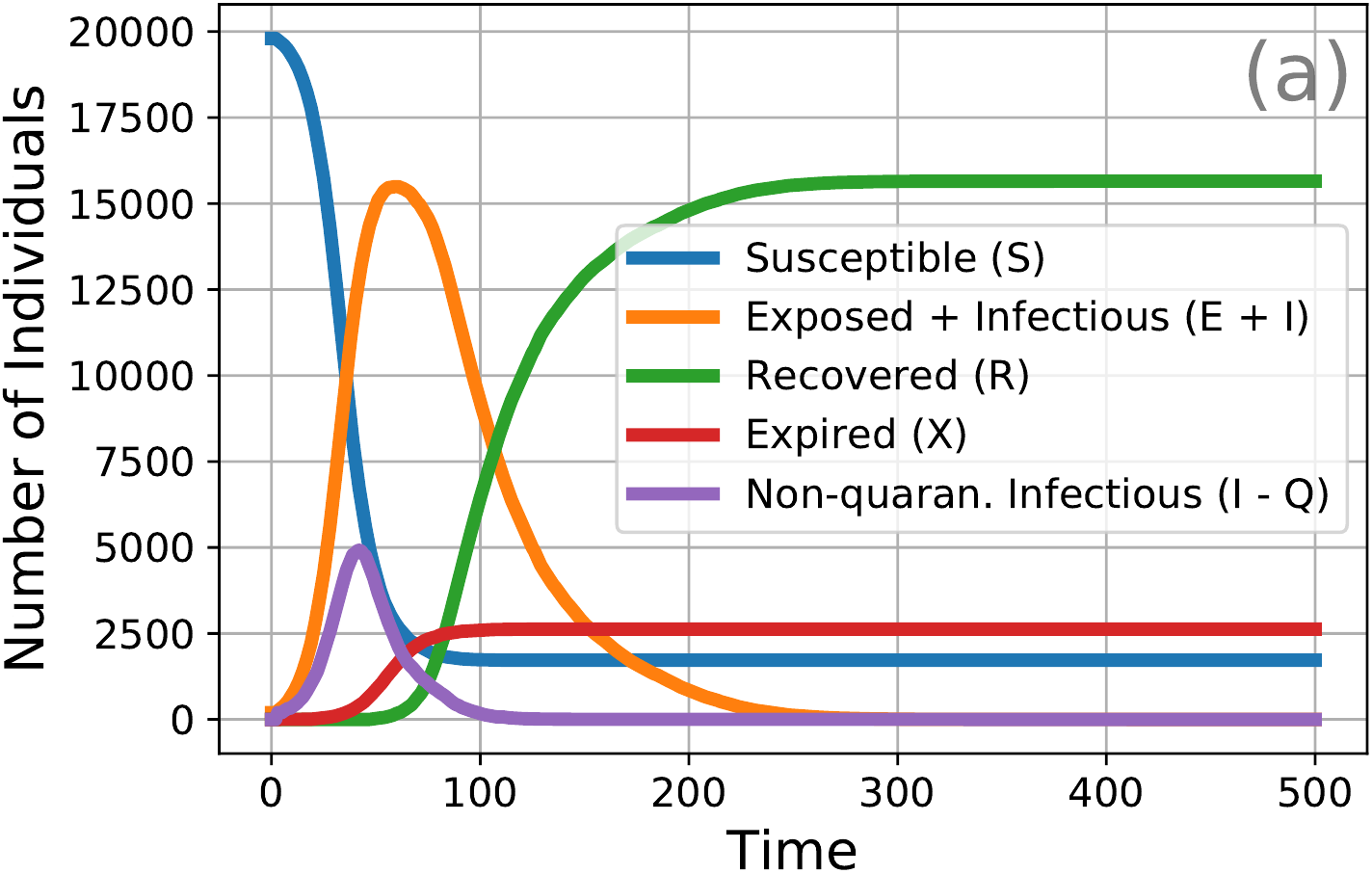}\hspace*{0.06\textwidth}
\includegraphics[width=0.41\columnwidth]{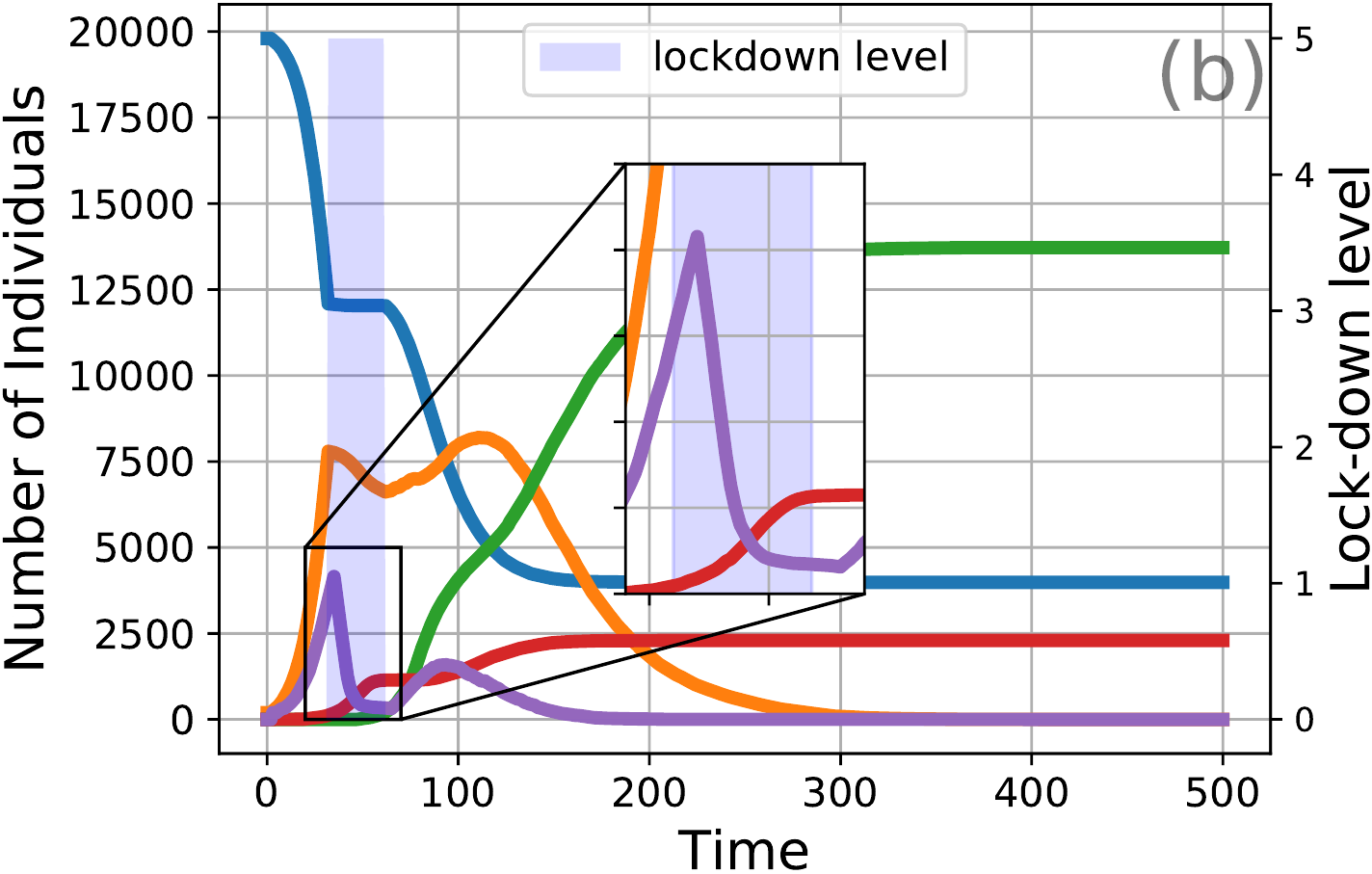}\hspace*{0.06\textwidth}\\%
\hspace*{0.06\textwidth}\includegraphics[width=0.41\columnwidth]{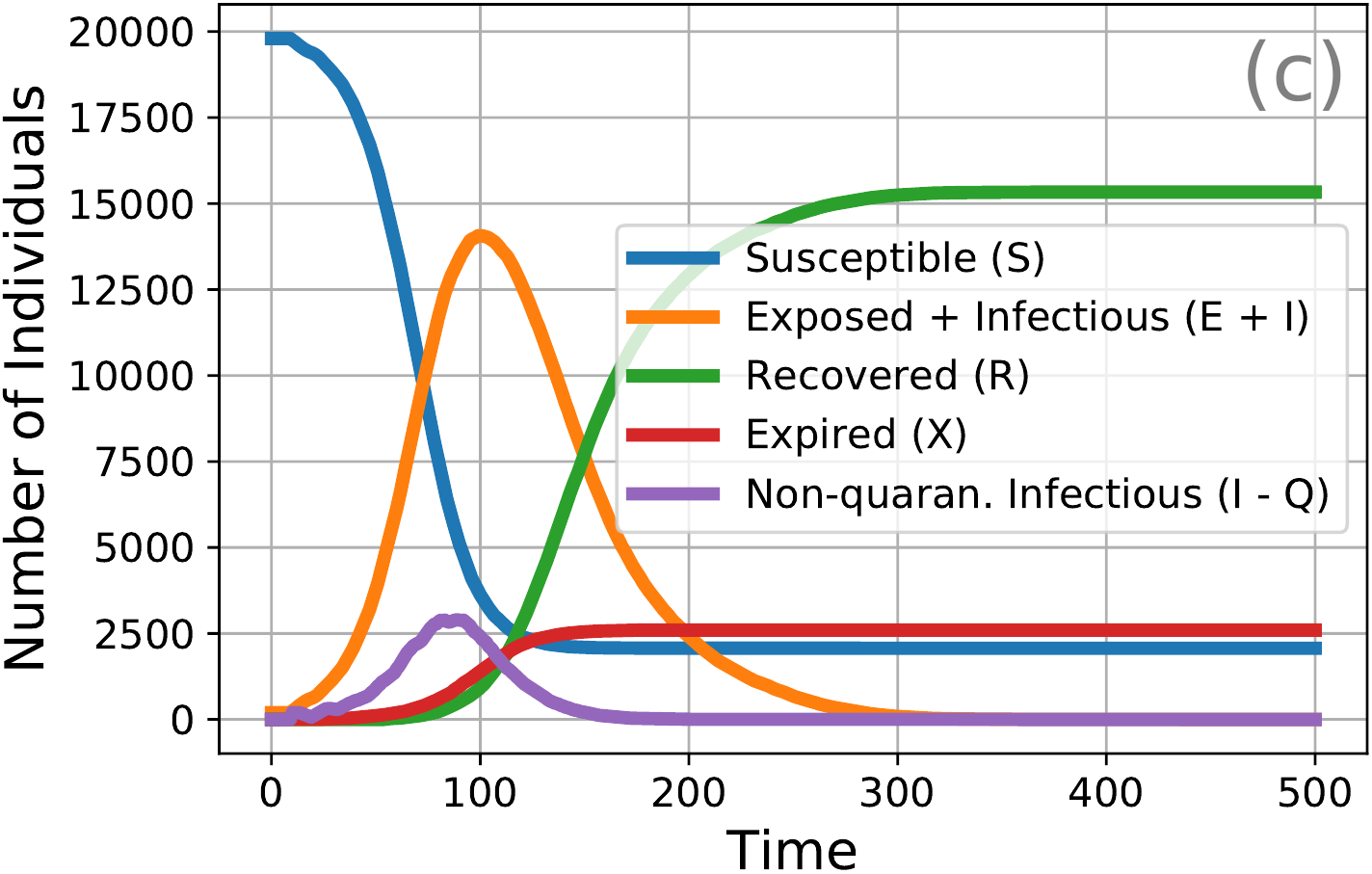}\hspace*{0.06\textwidth}
\includegraphics[width=0.41\columnwidth]{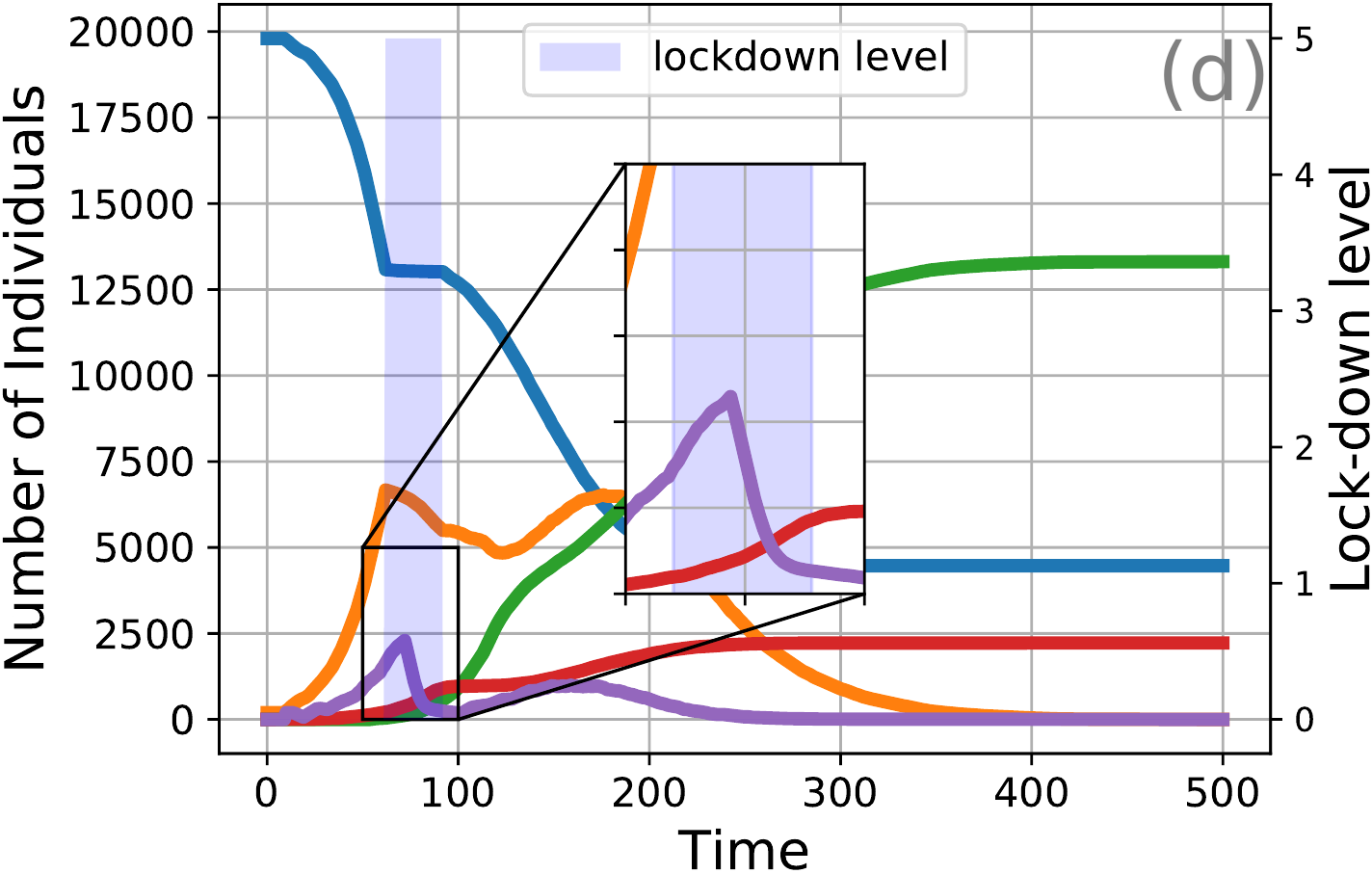}\hspace*{0.06\textwidth}%
\caption{\textbf{Using \myalgo to discover the optimal point at which to initiate a lock-down}. Figs~\ref{fig:single-phase-S}(a),(b) consider a virus strain with an incubation of 3 days whereas Figs~\ref{fig:single-phase-S}(c),(d) consider a strain with a 10 day incubation period. The optimal initiation point of a lock-down depends strongly on viral characteristics e.g. incubation period. \myalgo is able to discover a near-optimal initiation point in both cases within very few iterations (see Fig~\ref{fig:conv-rate}. See Sec~\ref{sec:single-phase-S} for details.}%
\label{fig:single-phase-S}
\end{figure}

\begin{figure}%
\hspace*{0.06\textwidth}\includegraphics[width=0.41\columnwidth]{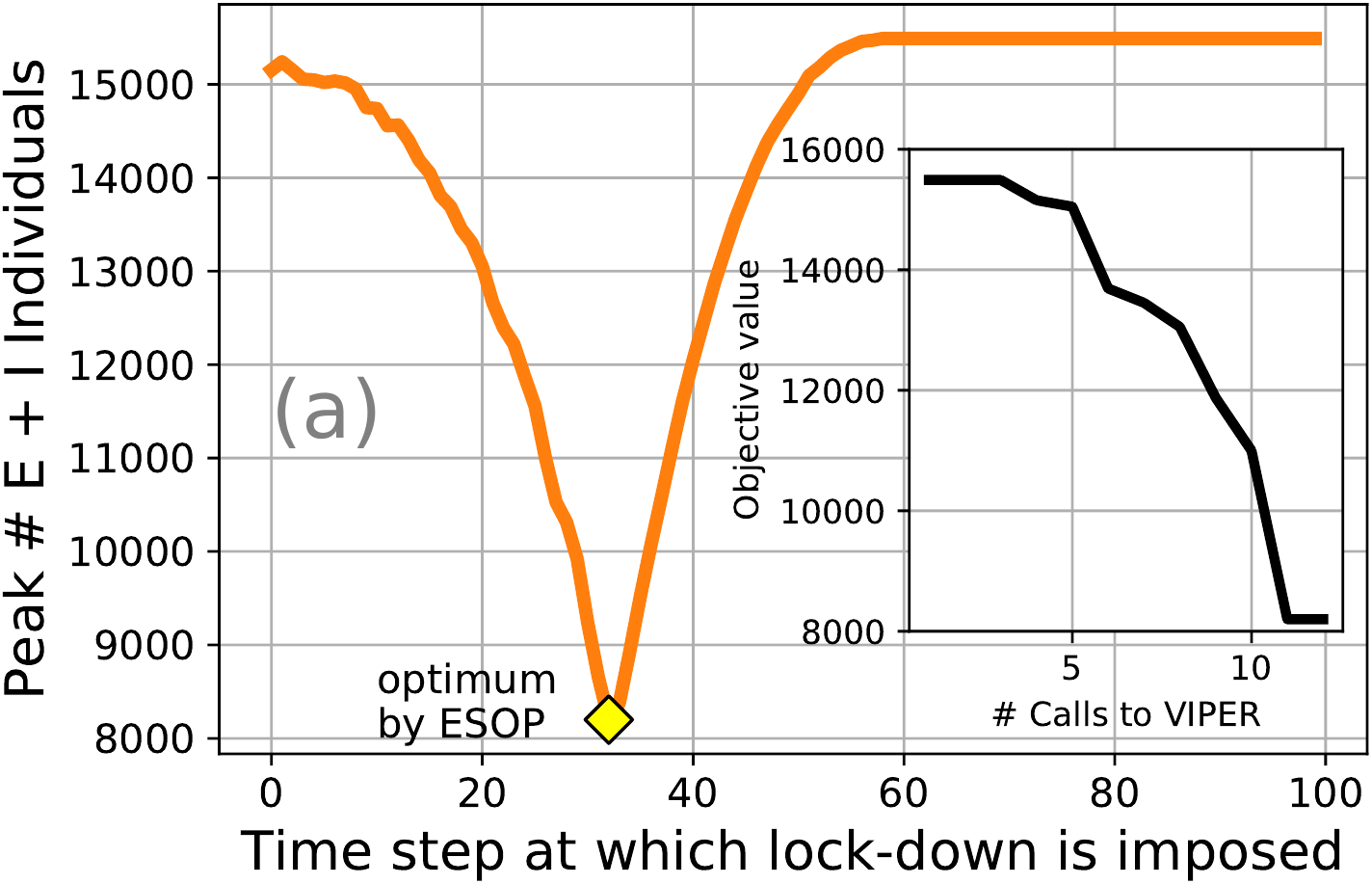}\hspace*{0.06\textwidth}%
\includegraphics[width=0.41\columnwidth]{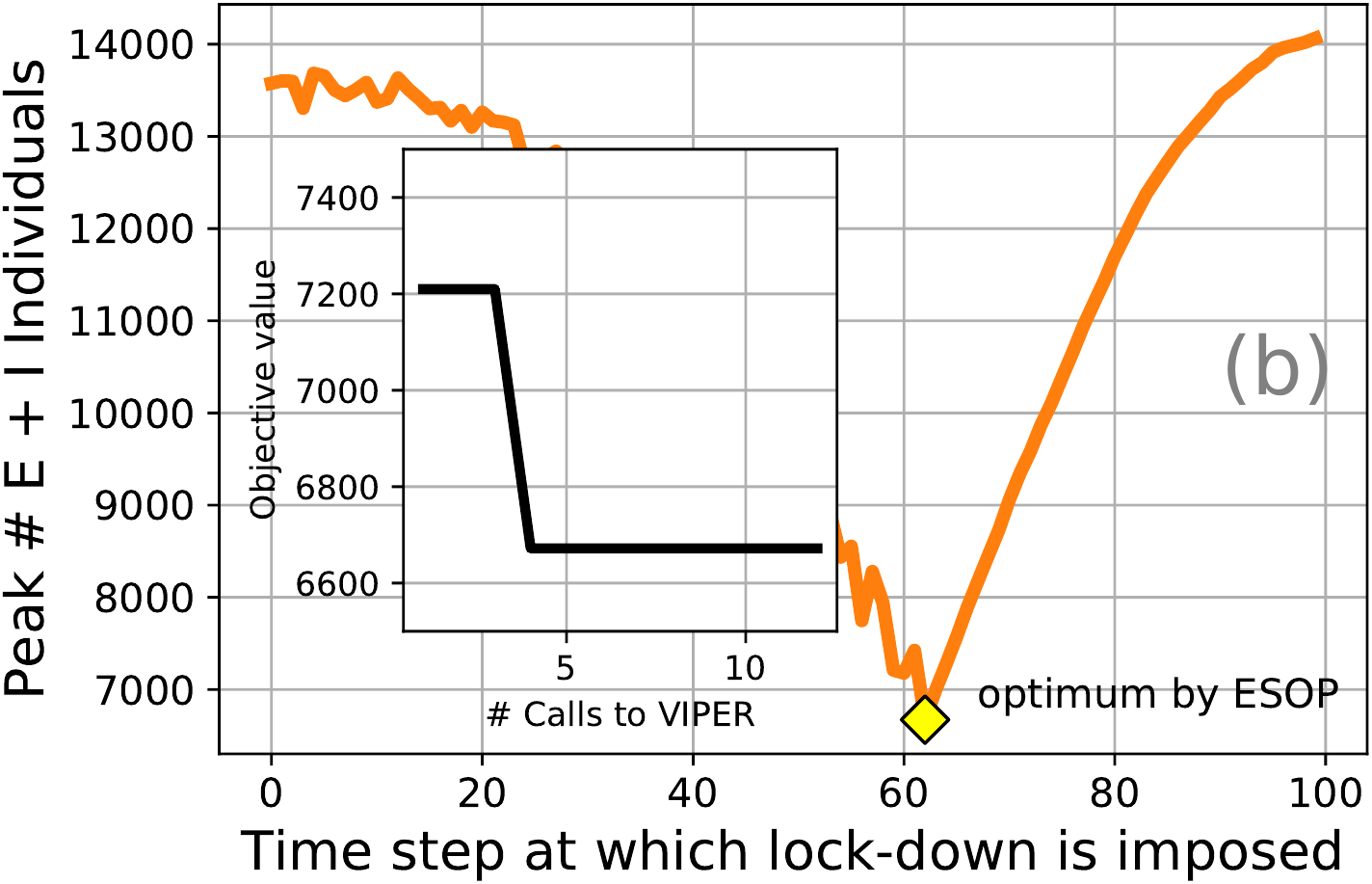}\hspace*{0.06\textwidth}%
\caption{\textbf{Convergence rates offered by \myalgo}. Figs~\ref{fig:conv-rate}(a),(b) show (see black curve) \myalgo's convergence rate in the 3 and 10 day incubation period cases in Fig~\ref{fig:single-phase-S}. The optimal initiation point varies significantly with the incubation period (33 vs 62 days). See~\ref{sec:conv-rate} for details.}%
\label{fig:conv-rate}%
\end{figure}

\section{Experimental Case Studies}
\label{sec:exps}
We present case studies with \mymodel simulating an in-silico population of 20000 individuals (\myalgo scales to larger populations too). The default attributes settings in \mymodel are given in Tab~\ref{tab:models}. Any modifications to these are mentioned below.

\subsection{Choice of $f_\epi$ and $f_\eco$}
A lock-down is represented as a 3-D vector $\vx = (i,p,l)$ of its initiation point ($i$), period ($p$) and level ($l$). Our objective is to minimize $f_\epi + f_\eco$ where $f_\epi(\vx)$ is the peak of the E+I curve. We use $f_\eco(\vx) = \frac l5 \cdot p \cdot \frac N{1000}$ to estimate job losses due to the lock-down assuming that a level-$l$ lock-down forces an $\frac l{50}$ percent of the population of $N = 20000$ into unemployment each day for $p$ days. Note that most natural definitions of $f_\epi$ and $f_\eco$ would conflict with each other since $f_\epi$ would promote aggressive and sustained lock-downs whereas $f_\eco$ would oppose them.

The notion of $f_\eco(\cdot)$ used above is merely demonstrative and \myalgo can instead use a more realistic definition of $f_\eco$ that might be a non-linear function of $\vx$, by asking an economic simulator, just as it asks values of $f_\epi$ from an epidemiological simulator such as \mymodel. Also, defining the objective as an additive sum $f_\epi + f_\eco$ of the two functions is not necessary essential for \myalgo to function and users may instead prefer other formulations e.g. $f_\eco^\alpha \cdot f_\epi^\beta$ for some $\alpha, \beta > 0$, etc.

All that \myalgo requires is black-box access to values of the objective function, however it may be defined. As Fig~\ref{fig:conv-rate} indicates, \myalgo is capable of optimizing highly non-linear functions as well. However, the algorithm would naturally require more iterations if the objective becomes highly convoluted and sensitive to changes in the input (see Sec~\ref{sec:conc} for a brief discussion).

\subsection{Using \myalgo to Find the Optimal Initiation Point of a Lock-down}
\label{sec:single-phase-S}
Fig~\ref{fig:single-phase-S} considers a simple case where we have decided to impose a 30 time step lock down at level 5 but are unsure when to initiate the lock down for optimal effect. The objective here is to minimize $f_\epi$ alone and $f_\eco$ is not considered since we have already decided the duration and intensity of the lock-down in this case and hence resigned to a predictable economic outcome. For any initiation point $i \in \bN$, let $f_\epi(i)$ be the largest number of individuals in E and I states at any given point of time (the so called \emph{peak} of the curve) if a level 5 lock-down is initiated at $t = i$.

Fig~\ref{fig:single-phase-S}(a) shows the daily count of individuals in various categories if no suppression is used. Note the large number of non-quarantined yet infectious individuals (I-Q) who are responsible for disease spread. The number of infected (E+I) individuals peaks at around 15500 at $t = 58$. For Fig~\ref{fig:single-phase-S}(b), \myalgo was asked to suggest when to start a 30 day lock-down at level 5. It suggested starting at $t = 33$ which brings the peak down to 8200 cases, a reduction of 47\%. Figs~\ref{fig:single-phase-S}(c), (d) show similar results but for a viral strain that has an incubation of 10 days instead of 3 days.

The results show that the optimal initiation point depends significantly on disease characteristics, e.g., incubation period of the virus. Nevertheless, \myalgo discovers a near-optimal solution, offering far superior health outcomes compared to a no-lock-down scenario. Note that the number of non-quarantined infectious individuals continues to rise (see Fig~\ref{fig:single-phase-S}(b), (d) insets) even after imposition of lock-down due to the incubation period of the disease.

\subsection{Convergence Rate of \myalgo}
\label{sec:conv-rate}
The topic of how fast do Bayesian optimization routines converge to the optimal solution is a subject of intense study but one beyond the scope of this paper. Under appropriate assumptions, Bayesian optimization routines, within $T$ queries to the underlying function (e.g., as \myalgo queries $f_\epi$ via \mymodel), are able to offer a solution that is only $\epsilon_T$ worse than the optimal solution. The sub-optimality $\epsilon_T$ goes down with the number of queries $T$ at a rate $\frac1{T^\alpha}$ where $\alpha > 0$ is a constant that depends on the problem setting and the \emph{kernel} used by the Bayesian optimization procedure (all experiments with \myalgo use the Matern kernel). We refer the reader to \citep{SrinivasKKS2010} for technical details of these convergence results.

Figs~\ref{fig:conv-rate}(a), (b) show that \myalgo achieved the globally optimal initiation point for the two cases considered in Fig~\ref{fig:single-phase-S} within just 12 and 4 calls to \mymodel (see inset figures). To plot the orange curves in Fig~\ref{fig:conv-rate}, \mymodel was queried with all $i \in [0,100]$ to explicitly reveal the globally optimum. This was done so that we could verify that \myalgo does indeed reach the global optimum. However, \myalgo itself requires far fewer, e.g. 12 or 4 calls to \mymodel to reach the global minimum.

\begin{figure}%
\hspace*{0.06\textwidth}\includegraphics[width=0.41\columnwidth]{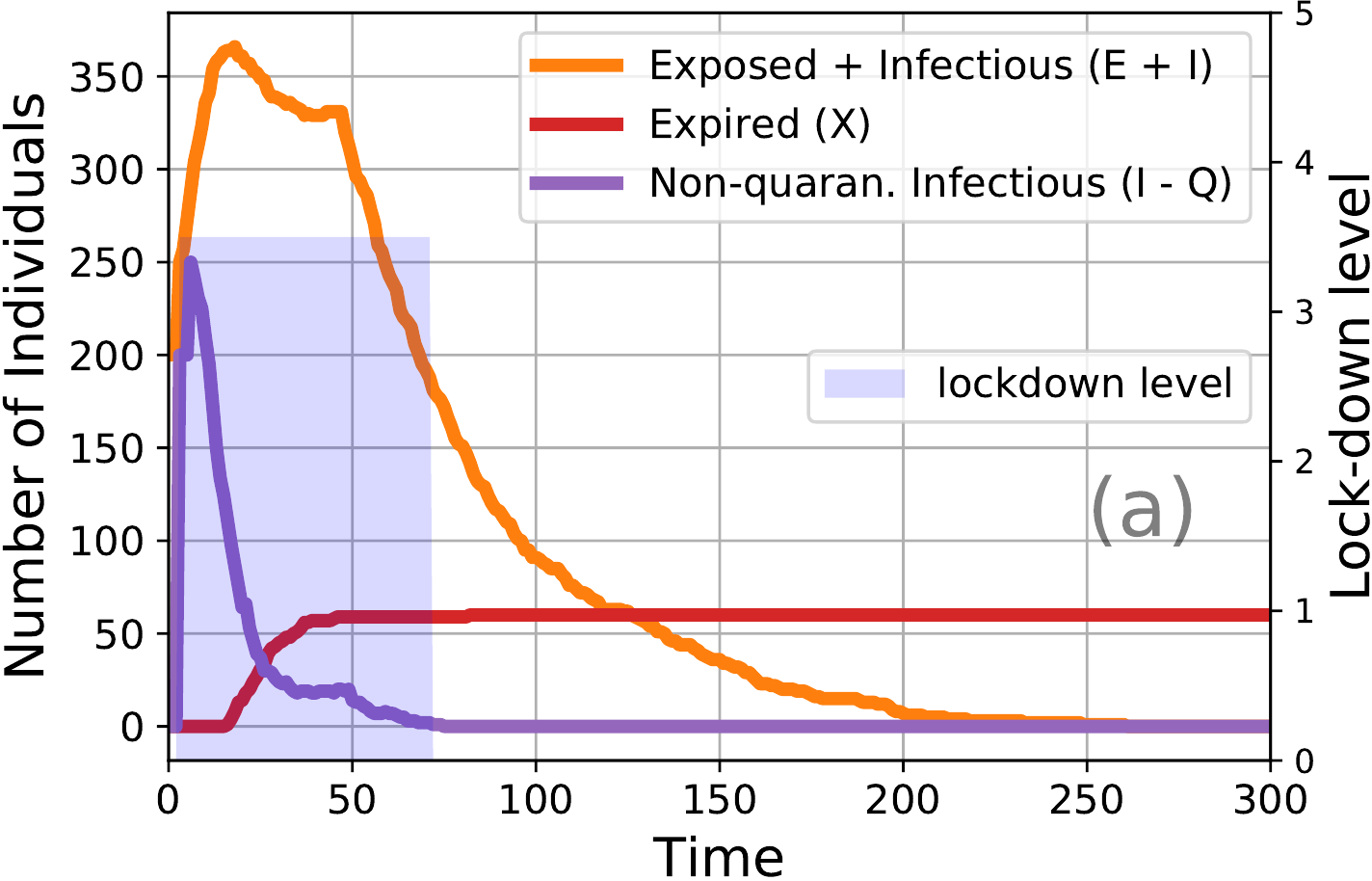}\hspace*{0.06\textwidth}%
\includegraphics[width=0.41\columnwidth]{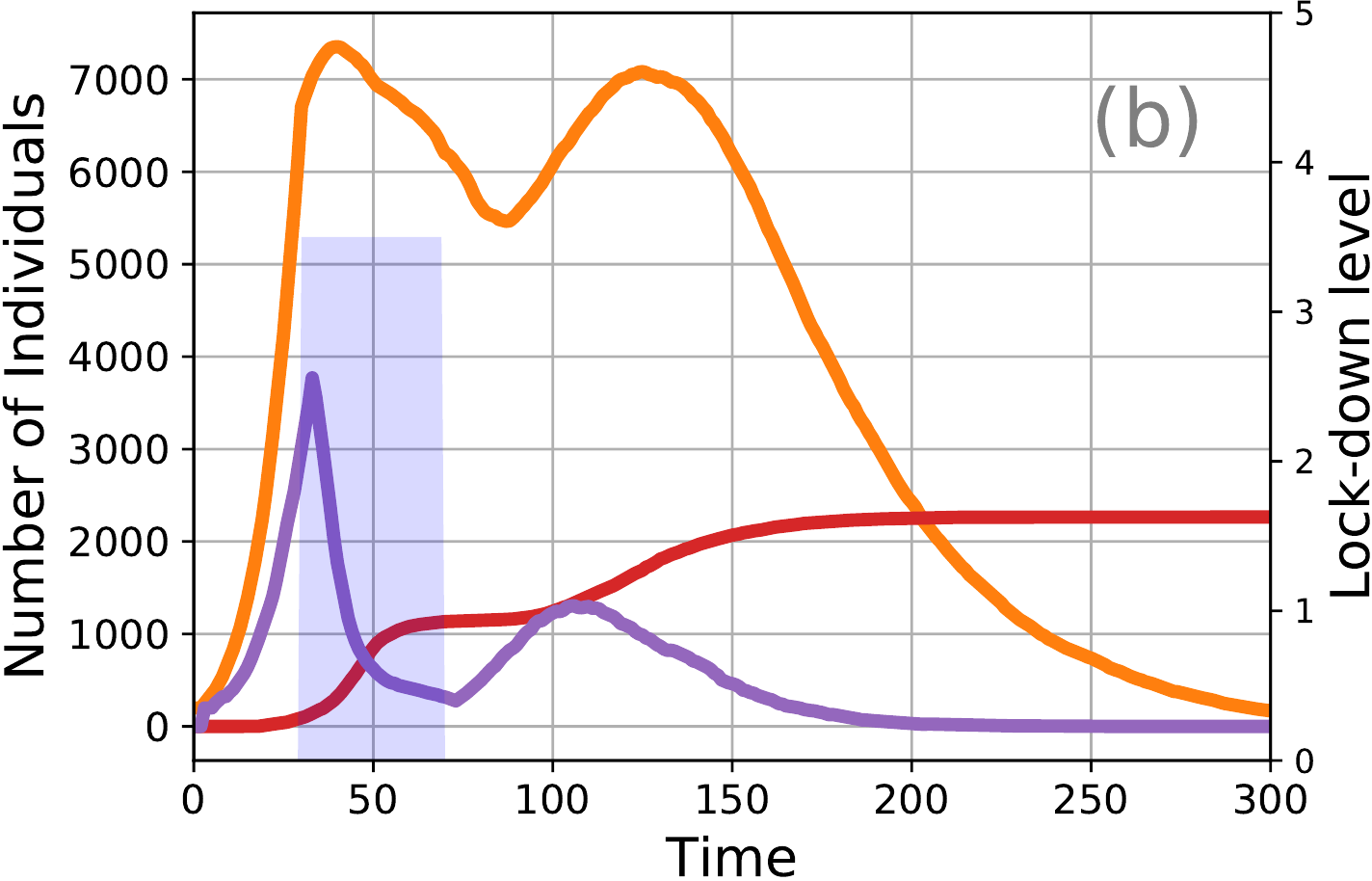}\hspace*{0.06\textwidth}\\%
\hspace*{0.06\textwidth}\includegraphics[width=0.41\columnwidth]{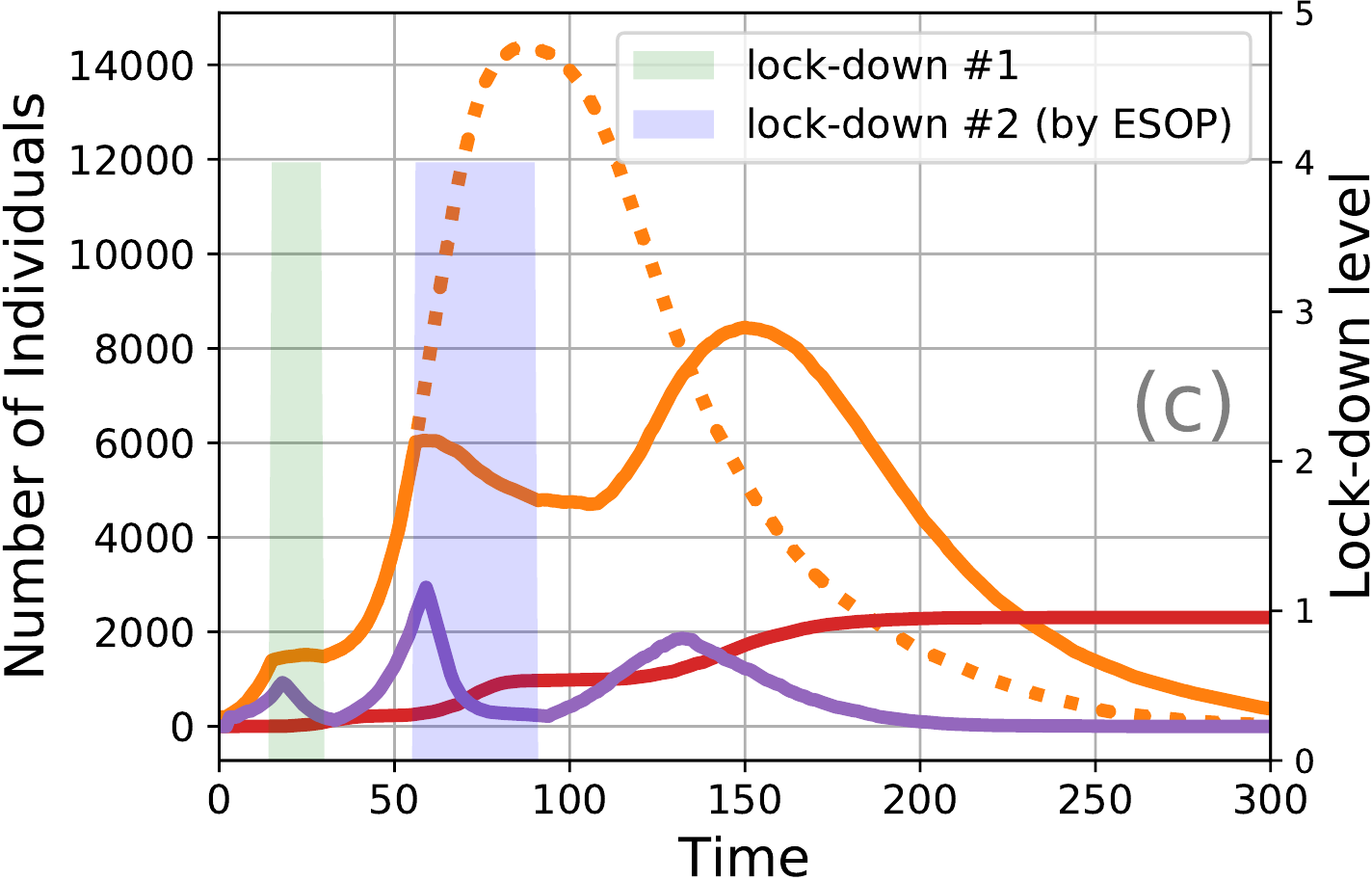}\hspace*{0.06\textwidth}%
\includegraphics[width=0.41\columnwidth]{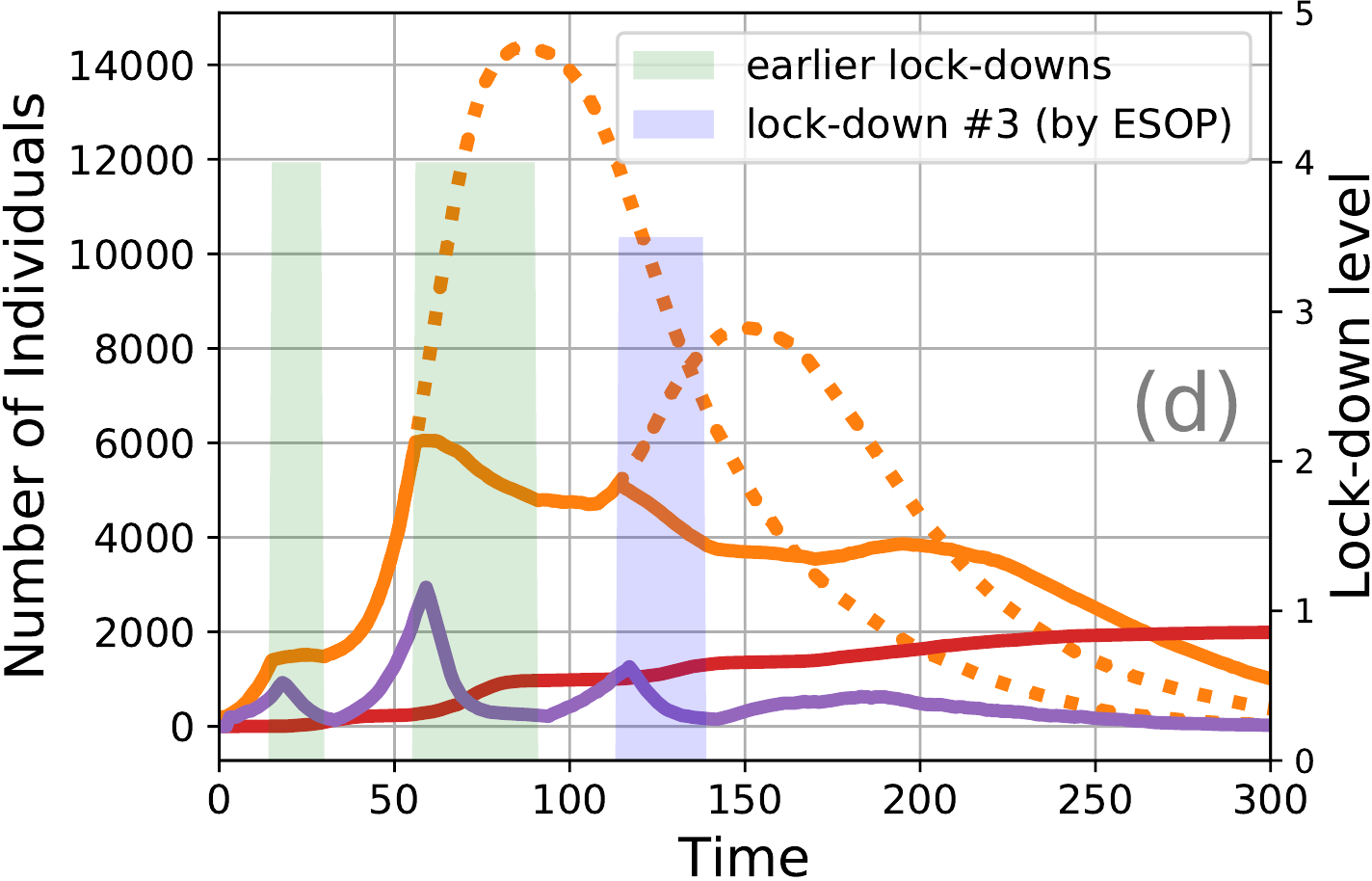}\hspace*{0.06\textwidth}%
\caption{\textbf{Optimizing (multi-phase) lock-downs under constraints}. See Sec~\ref{sec:single-phase-SPL} for details. \myalgo is able to shift from a containment strategy (e.g. in Fig~\ref{fig:single-phase-SPL}(a)) to a mitigation strategy (e.g. in Fig~\ref{fig:single-phase-SPL}(b)) depending on the constraints. When multiple spikes are inevitable due to constraints on the lock-down (e.g. upper limits on the duration of the lock-downs), \myalgo balances the heights of the spikes to ensure that infections are evenly distributed among them.}%
\label{fig:single-phase-SPL}
\end{figure}

\subsection{Using \myalgo to Optimize (Multi-phase) Lock-downs under Constraints}
\label{sec:single-phase-SPL}
Administrative and logistical considerations put constraints on lock-downs, e.g., on when a lock-down can be initiated or, on how long it can last. Another constraint might be taking into account previously applied lock-downs.  We demonstrate how \myalgo can be used to design lock-down schedules that are optimal among those that satisfy a given set of constraints. Fig~\ref{fig:single-phase-SPL} considers finding the optimal initiation, period and level of a lock-down, subject to constraints.

\paragraph{Single-phase Lock-downs under Constraints.} For Fig~\ref{fig:single-phase-SPL}(a), \myalgo was asked to suggest a lock-down (no constraints on duration etc). It suggested one at level 3.5 starting $t = 3$, lasting 69 steps, resulting in a peak of 303 infections (much smaller than the peaks in Fig~\ref{fig:single-phase-S}) and 1092 predicted cases of unemployment. We note that $f_\eco$ prevents the lock-down from going on indefinitely. The suggested lock-down seems to adopt a containment strategy, initiating a moderate-level suppression very early on to deplete the pool of infectious individuals. Note that the pool of non-quarantined infectious individuals is indeed exhausted by the time the suggested lock-down is over, preventing a second wave of infections. For Fig~\ref{fig:single-phase-SPL}(b), \myalgo was forced to suggest a lock-down starting no earlier than $t = 12$ and lasting no longer than 40 days. Since infections are already rampant by $t = 12$, \myalgo instead adopts a mitigation strategy of starting a lock-down at $t = 30$ for 40 steps at level 3.5, causing a peak of 7353 infections and 560 cases of unemployment. Notice that the lock-down is strategically delayed so that the second wave does not have a higher peak, thus balancing the two peaks indeed as dictated by $f_\epi$.

\paragraph{Multi-phase Lock-downs.} For Fig~\ref{fig:single-phase-SPL}(c), \myalgo was given a situation where an earlier lock-down (green shading) had already taken place but was ineffective and left alone, would have caused a massive second wave with a peak of 14384 (dotted orange curve). \myalgo was asked to suggest a new lock-down that starts no earlier than 10 days after the previous lock-down ended and lasting no longer than 40 days. \myalgo suggested a second lock-down starting at $t = 56$ lasting 35 steps at level 4 which brings the peak down to 8447 (a reduction of 40\%) and causing 560 additional cases of unemployment. Fig~\ref{fig:single-phase-SPL}(d) considers a scenario where the policy makes are still dissatisfied with the outcomes, and request a third lock-down starting no earlier than 10 days after the second lock-down ended and lasting no longer than 40 days. \myalgo suggests a third milder lock-down at level 3.5 starting at $t = 114$ and lasting 25 days. This brings the peak to a much lower number (6052 i.e. a further 28\% reduction) and 350 additional cases of unemployment.

\subsection{Can aggressive quarantining permit \myalgo to offer less severe lock-downs?}
\label{sec:qrn}
\begin{figure}%
\hspace*{0.06\textwidth}\includegraphics[width=0.41\columnwidth]{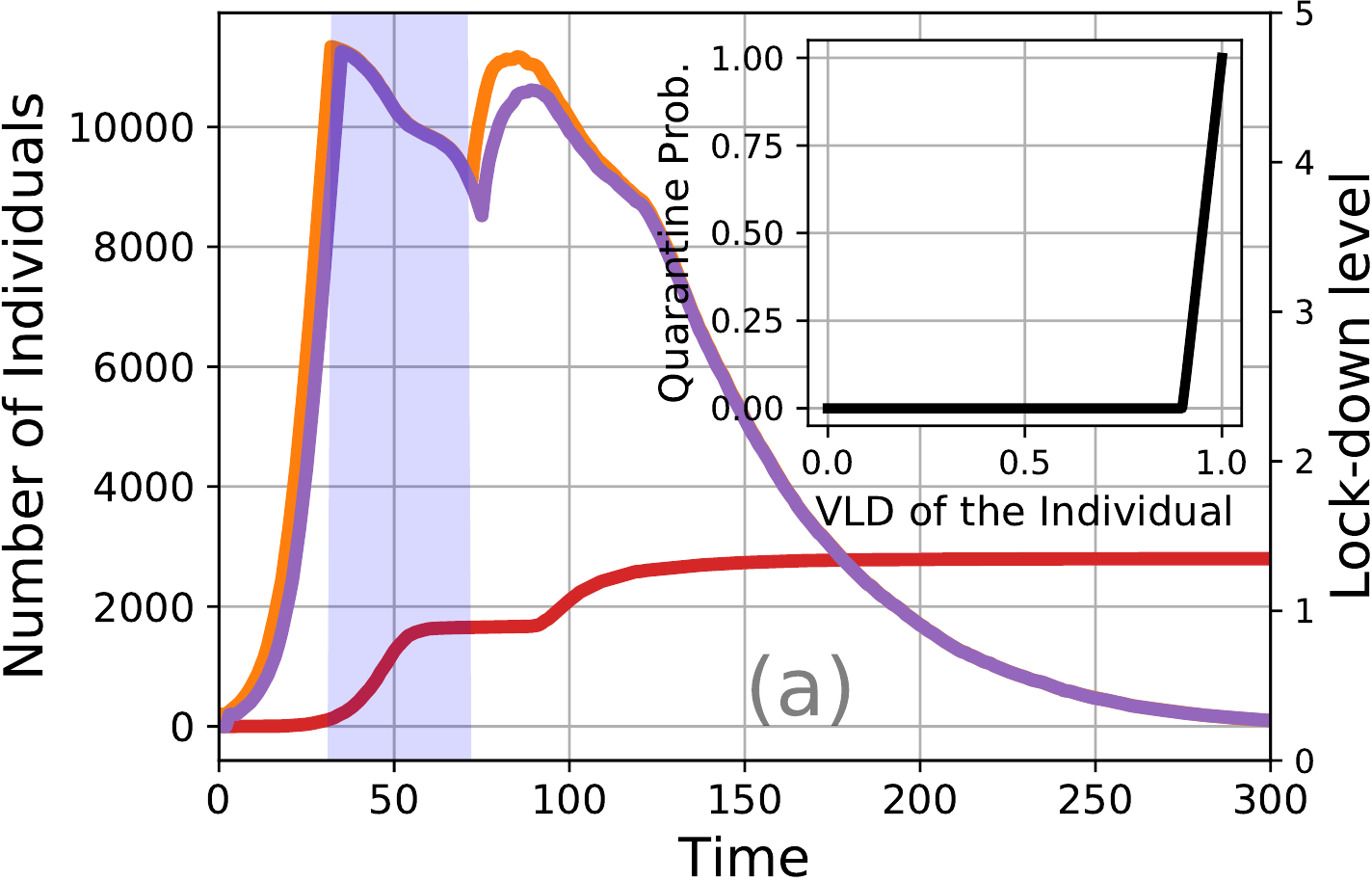}\hspace*{0.06\textwidth}%
\includegraphics[width=0.41\columnwidth]{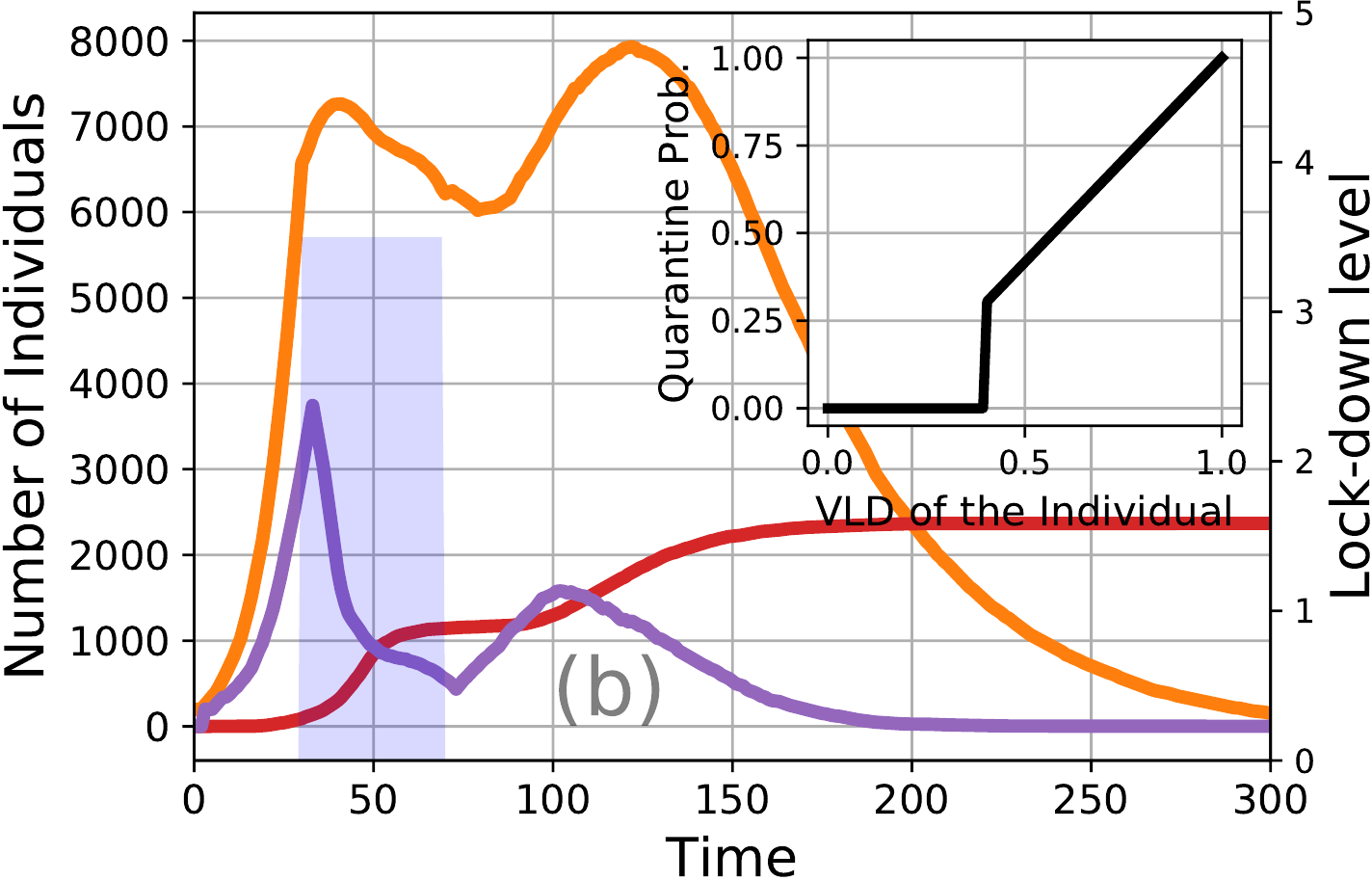}\hspace*{0.06\textwidth}\\%
\hspace*{0.06\textwidth}\includegraphics[width=0.41\columnwidth]{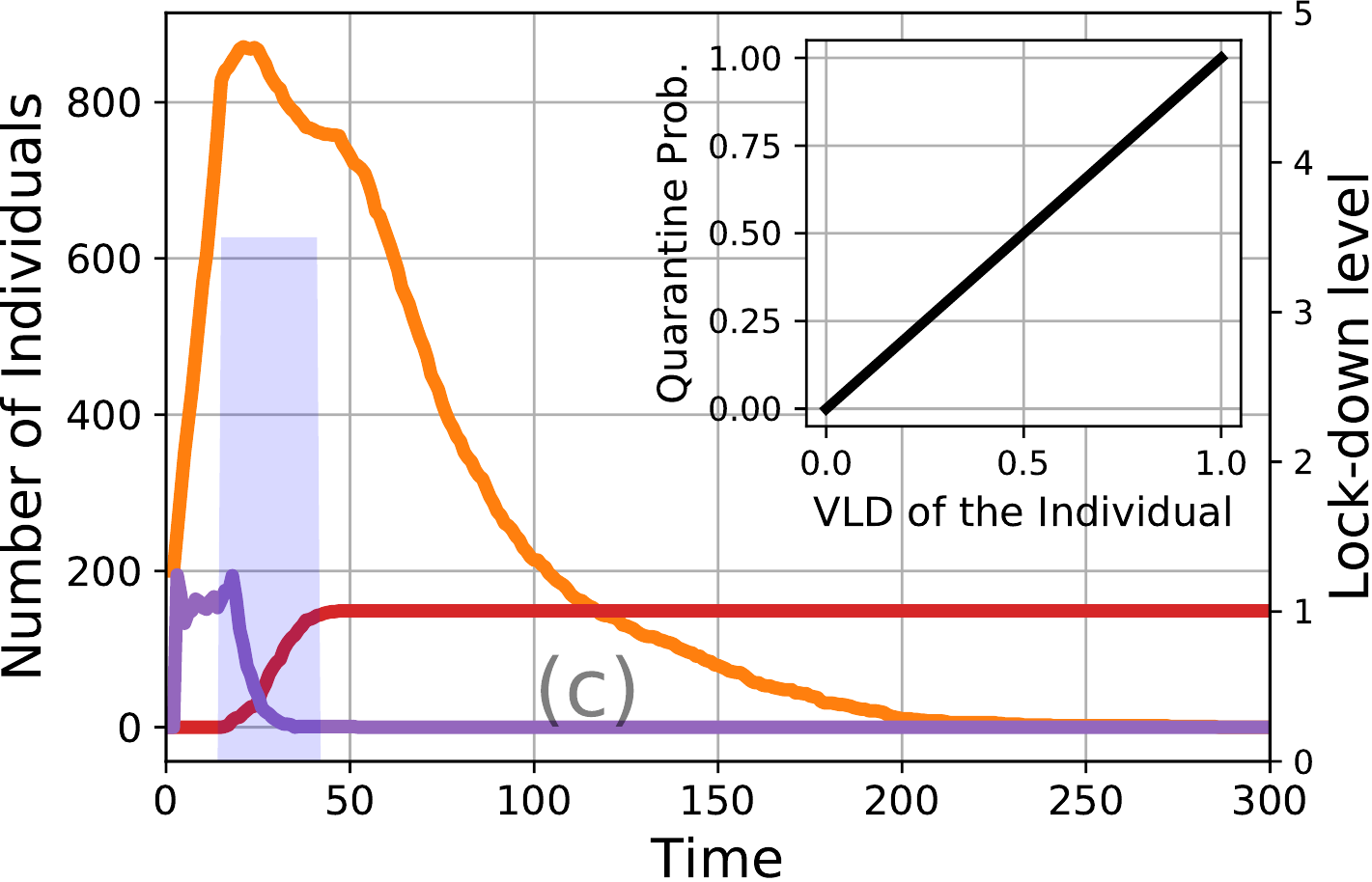}\hspace*{0.06\textwidth}%
\includegraphics[width=0.41\columnwidth]{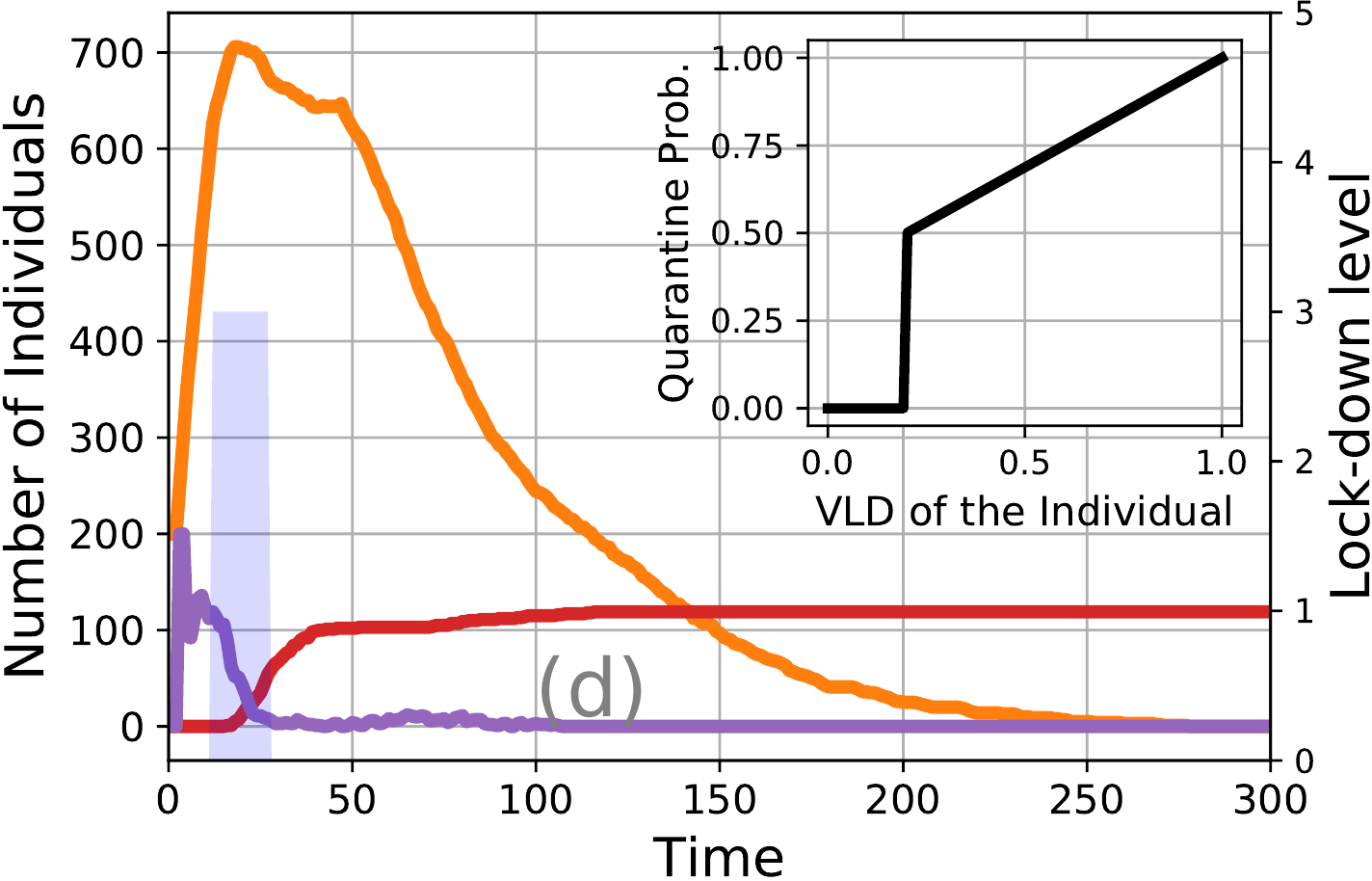}\hspace*{0.06\textwidth}%
\caption{\textbf{Effect of quarantine policy on \myalgo's ability to offer desirable outcomes}. The area under the black curve in the insets roughly measures the aggressiveness of the quarantining policy. See Sec~\ref{sec:qrn} for details. \myalgo offers better outcomes, both epidemiologically as well as economically, and that too with milder lock-downs, if strong quarantining is applied.}%
\label{fig:single-phase-SPL-QRN}
\end{figure}

Intuitively, if aggressive quarantining is applied, then it should be possible to avert an epidemic with milder lock-downs. Fig~\ref{fig:single-phase-SPL-QRN} verifies this claim by considering at the situation in Fig~\ref{fig:single-phase-SPL}(b) (i.e. starting a lock-down no earlier than $t = 12$ days and lasting no longer than 40 days), but with varying quarantine aggressiveness. As Tab~\ref{tab:models} and Sec~\ref{sec:problem} explain, individuals with viral loads over a threshold QTH get quarantined with a certain probability profile. Using greater screening and public awareness, this profile can be altered.

Fig~\ref{fig:single-phase-SPL-QRN}(a) considers sluggish quarantining with QTH = 0.9 and BQP = 0.0 (the quarantining probability profile is shown in Fig~\ref{fig:single-phase-SPL-QRN}(a) as an inset). Almost no individuals get quarantined and despite its best efforts, \myalgo is only able to offer 11338 peak infections and 800 unemployment cases using a lock-down starting at $t = 32$ (level 5, 40 steps). With stronger quarantining at QTH = 0.4 and BQP = 0.3, the situation improves in Fig~\ref{fig:single-phase-SPL-QRN}(b) where \myalgo offers fewer infections (7930) and job losses (560) using a lock-down at a reduced level of 3.5 (starting $t = 30$, 40 steps). With still stronger quarantining (QTH = 0.0, BQP = 0.0) in Fig~\ref{fig:single-phase-SPL-QRN}(c), \myalgo offers a peak of 871 and 378 job losses using a lock-down lasting fewer (27) steps (starting $t = 15$, level 3.5). Fig~\ref{fig:single-phase-SPL-QRN}(c) reports the outcomes with even stronger quarantining at QTH = 0.2 and BQP = 0.5 where \myalgo offers a peak of just 706 and 192 job losses using a lock-down lasting just 16 steps(level 3 starting $t = 12$).

\subsection{Variations in Demographics and Geographical Distribution of Population}
\label{sec:fitted}
We also consider location (X, Y), susceptibility (SUS) and resistance (RST) values (see Tab~\ref{tab:models}) which are not set uniformly but fitted to statistical data for India.

\noindent\textbf{Population Clusters.} Instead of distributing individuals uniformly in the box $[0,1]^2$, as we did earlier, we now distribute 34\% of the population into 4 Gaussian clusters with standard deviation 0.1 (to simulate crowded urban areas). India does have a similar proportion of urban population \citep{WB2018}. We distribute the rest 66\% of the population uniformly (to simulate scattered non-urban population).

\begin{figure}%
\hspace*{0.06\textwidth}\includegraphics[width=0.41\columnwidth]{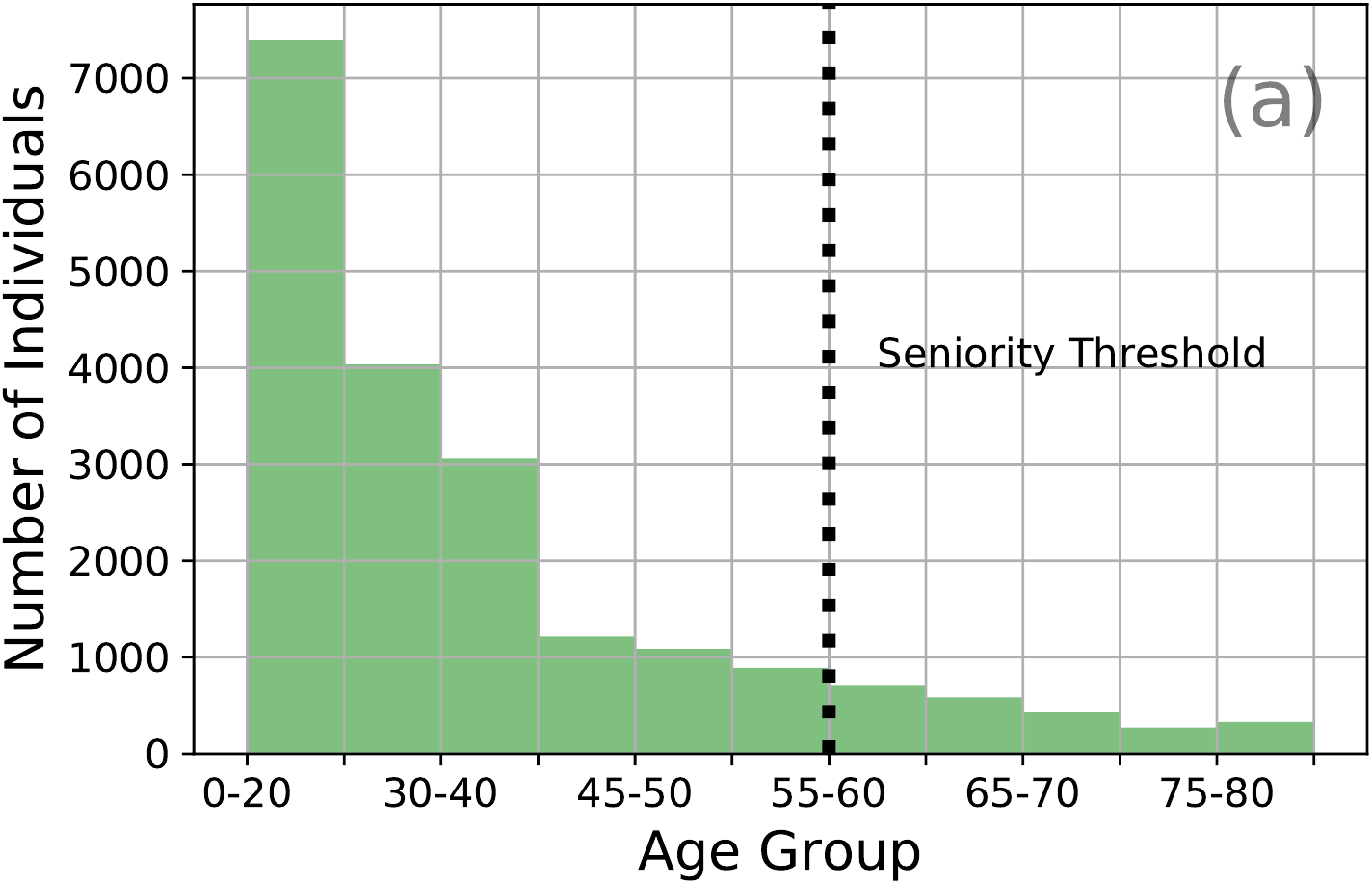}\hspace*{0.06\textwidth}%
\includegraphics[width=0.41\columnwidth]{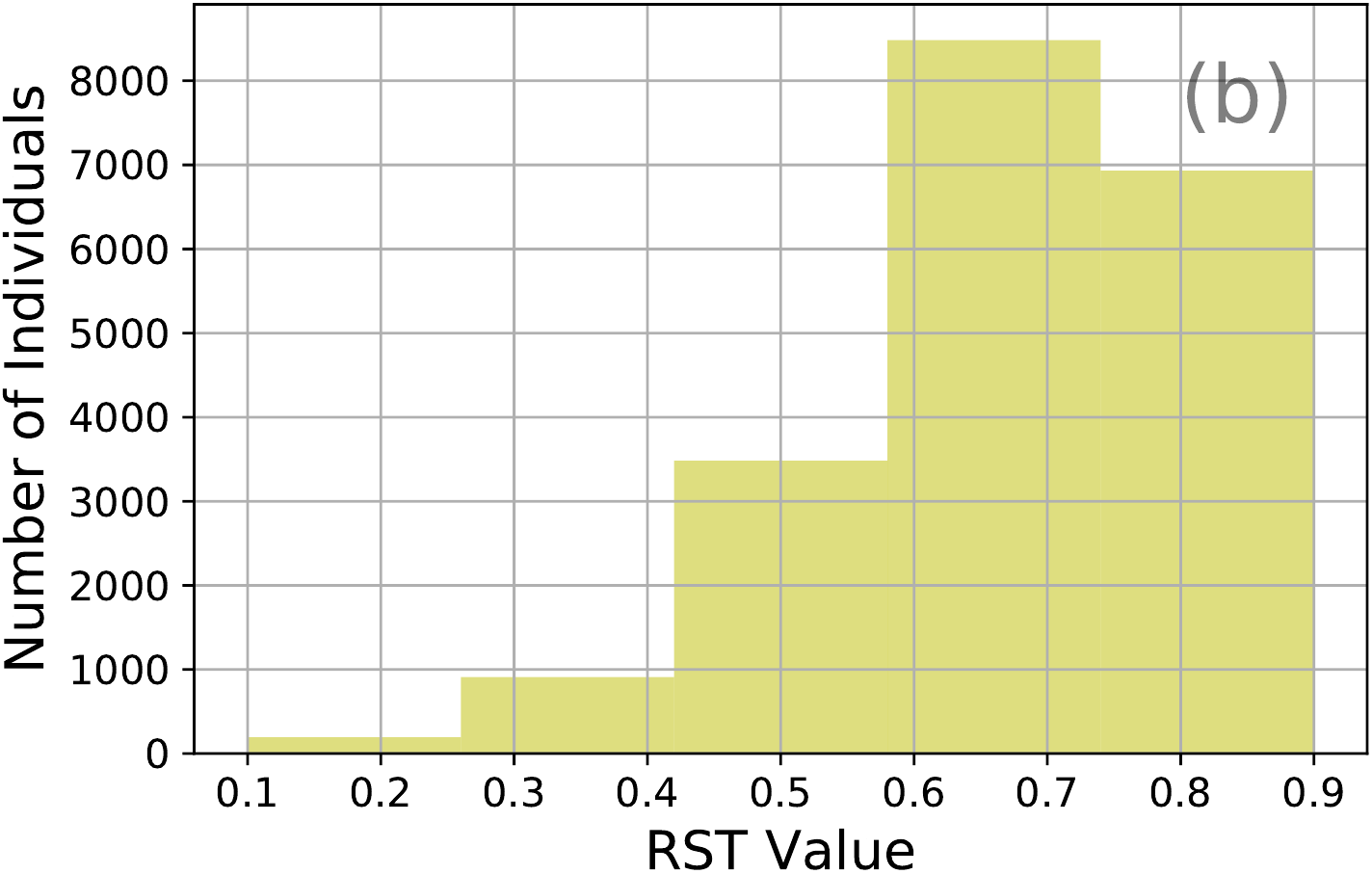}\hspace*{0.06\textwidth}
\caption{\textbf{Demographic data fitted to Indian statistics}. The demographic advantage of the Indian population is evident from both curves. Fig~\ref{fig:new-stats}(a) shows that the vast majority of the population do not cross the risk-factor forming seniority threshold whereas Fig~\ref{fig:new-stats}(b) shows that most individuals have high RST value i.e. resistance to infection, as calculated in Sec~\ref{sec:fitted}.}%
\label{fig:new-stats}%
\end{figure}

\begin{figure}%
\hspace*{0.06\textwidth}\includegraphics[width=0.41\columnwidth]{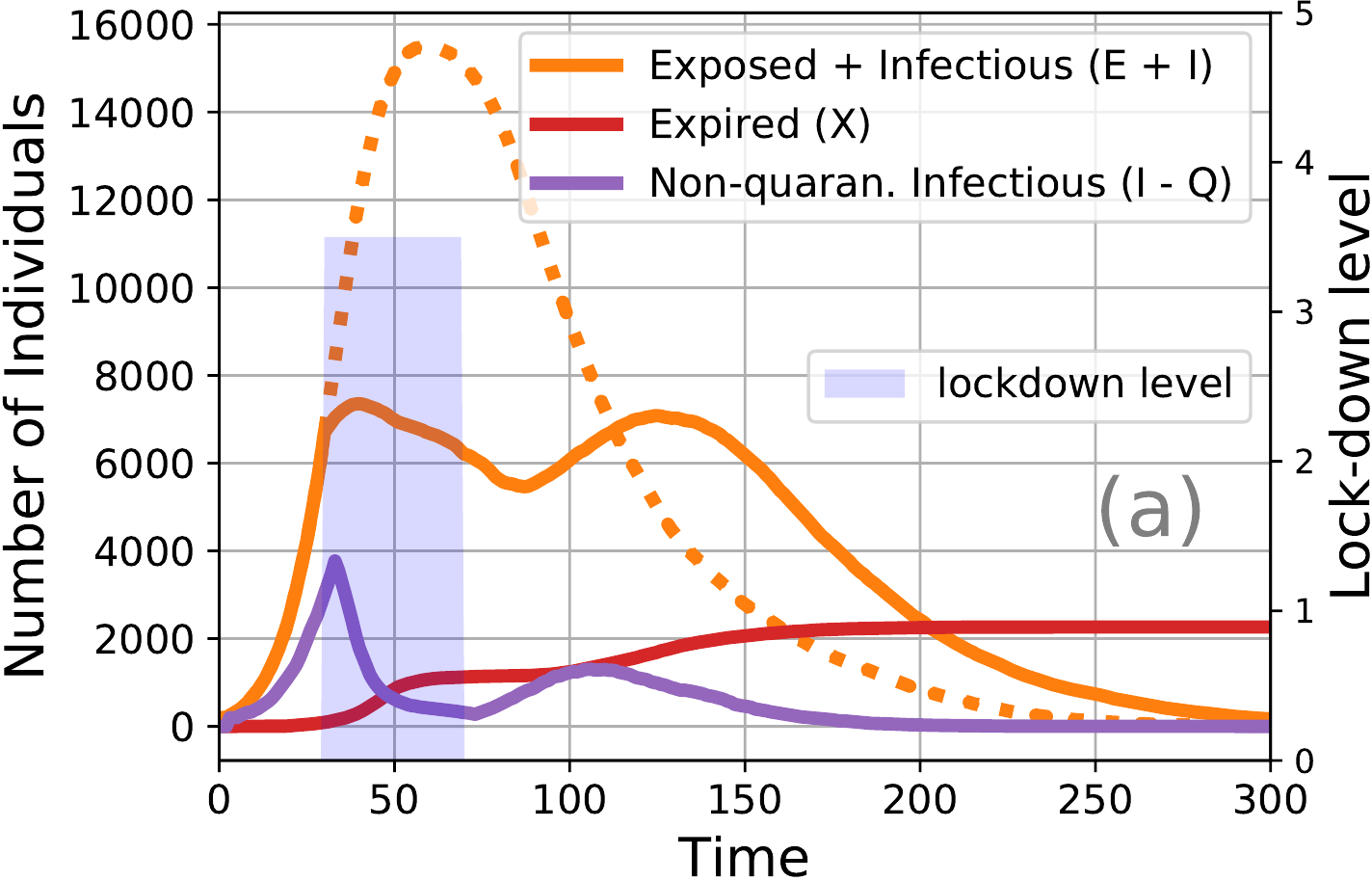}\hspace*{0.06\textwidth}%
\includegraphics[width=0.41\columnwidth]{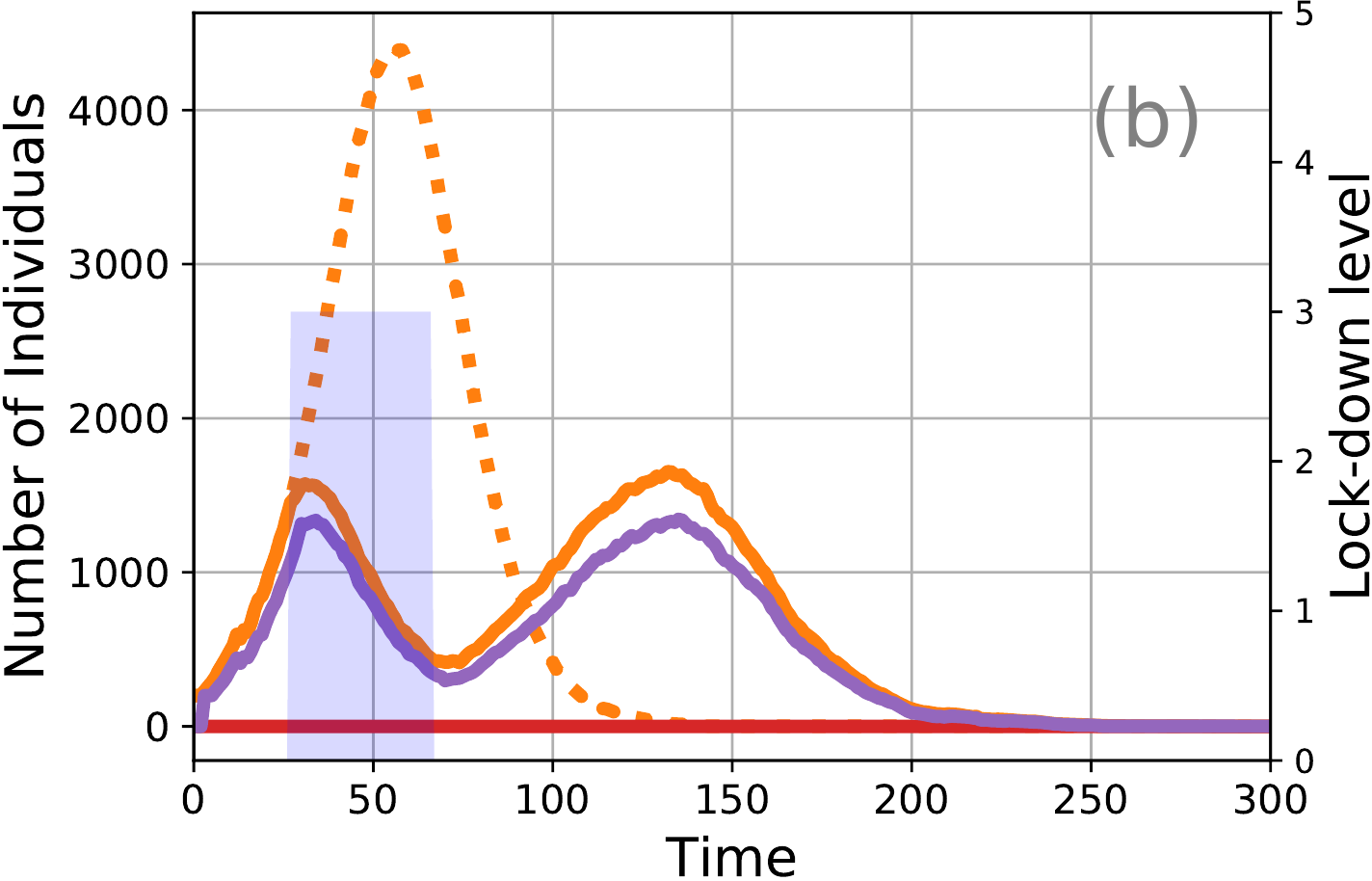}\hspace*{0.06\textwidth}\\%
\hspace*{0.06\textwidth}\includegraphics[width=0.41\columnwidth]{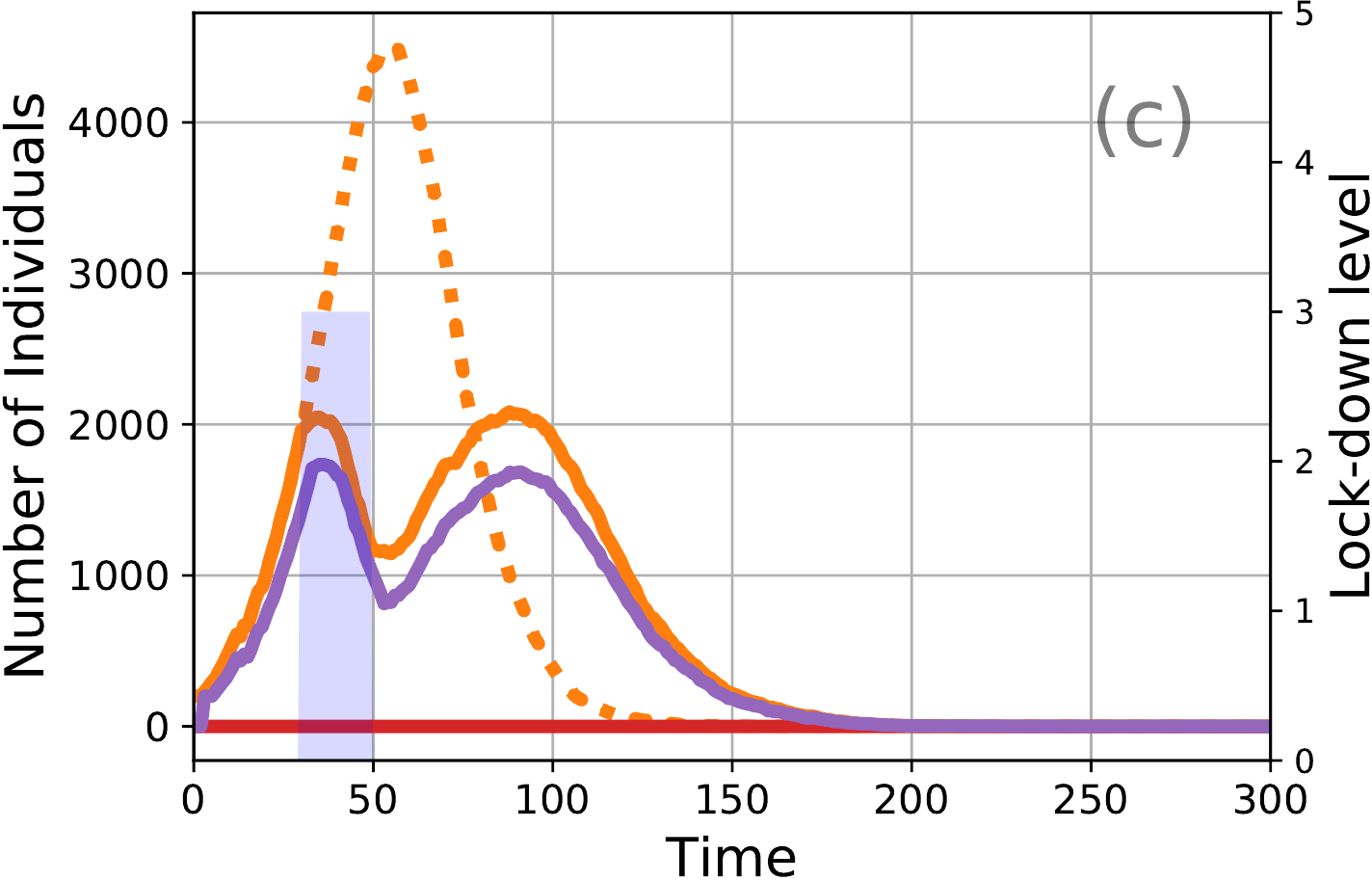}\hspace*{0.06\textwidth}%
\includegraphics[width=0.41\columnwidth]{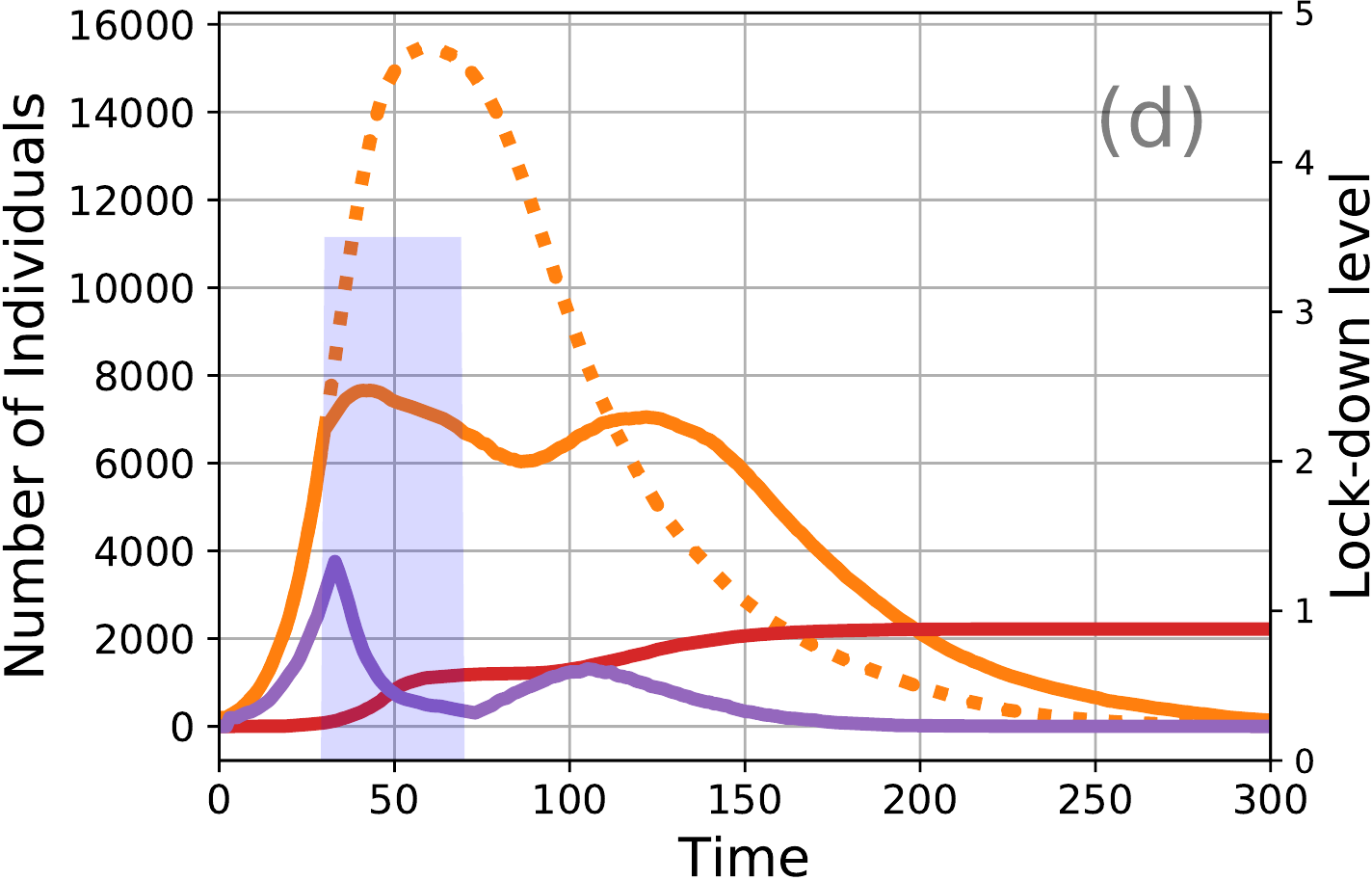}\hspace*{0.06\textwidth}%
\caption{\textbf{\myalgo with fitted demographic and location data}. See Sec~\ref{sec:new-results} for details. The dotted lines in all figures gives the infection curve had no lock-down been applied.}%
\label{fig:new-results}%
\end{figure}

\noindent\textbf{Fitted Demographic Data.} Instead of distributing SUS and RST values uniformly in $[0,1]$ as we did earlier, we now distribute these according to age and co-morbidity statistics for India. \citep[Table 1]{Joshi2020} reports that the four factors with highest risk score for CoViD-19 patients are age above 55 years, male gender, hypertension, and diabetes. We collected age-stratified statistics for all these factors from respectively \citep[Detailed Tables]{SRS2016}, \citep[Table 1]{Ramakrishnanetal2019} and \citep[Table 2]{DECODA2003}. After assigning age, gender, and co-morbidity values to alll individuals so as to fit the Indian statistics, individuals were assigned a point each for being male, over 55, and suffering from each of the two co-morbidities. Thus, an individual could clock up a maximum of 4 points. An individual's SUS value was then decided as $0.1 + 0.2 \times \text{ \#points }$. Their RST value was set to $\text{RST} = 1 - \text{SUS}$. Fig~\ref{fig:new-stats}(b) shows the resulting distributions.

\subsubsection{Results with Fitted Demographic Data and Population Clusters}
\label{sec:new-results}
Fig~\ref{fig:new-results}(a) is identical to Fig~\ref{fig:single-phase-SPL}(b) (where \myalgo was asked for a lock-down lasting no longer than 40 days and starting no earlier than $t = 12$) except that Fig~\ref{fig:new-results}(a) also shows (using a dotted orange line) what the infection curve would have been, had no lock-down been applied at all. Recall that with the lock-down suggested by \myalgo, we had a peak of 7353 infections and 560 cases of unemployment. Fig~\ref{fig:new-results}(b) presents the situation where we use population clusters as well as fitted demographic data. The improvement with \myalgo's new suggestion is extremely significant with a peak of just 1653 infections (more than $4\times$ reduction) and 480 cases of unemployment, that too, using a milder lock-down at level 3.0.

To ascertain whether this improvement was due to the demographic change or the population clusters, Figs~\ref{fig:new-results}(c),(d) present ablation studies. In Fig~\ref{fig:new-results}(c), only the demographic distributions of RST and SUS are fitted to Indian data but locations are kept uniformly distributed over the 2-D box $[0,1]^2$ as we were doing earlier. In Fig~\ref{fig:new-results}(d), RST and SUS values are uniformly distributed over $[0,1]$ as we were doing earlier, but 34\% of the population is clustered into ``cities''.

It is clear that changing the location distribution of the individuals alone worsens the outcome (Fig~\ref{fig:new-results}(d) reports a peak of 7657 that is higher than Fig~\ref{fig:new-results}(a)). This is to be expected since people are now crowded into cities . However, changing the demographics of the population to match that of India greatly improves the outcome (Fig~\ref{fig:new-results}(c) reports a peak of just 2083 and just 240 cases of unemployment). This is also to be expected since India has an extremely favorable age structure. Since applying both changes together, as in Fig~\ref{fig:new-results}(b), also presents a significant improvement over Fig~\ref{fig:new-results}(a), this seems to suggest that the demographic advantages more than overcome the disadvantage due to crowding in cities.

\section{Concluding Remarks}
\label{sec:conc}
Incorporating age stratification and climate into \mymodel and augmenting \myalgo to suggest ``personalized'' age- and climate-specific policies would be interesting. It would also be interesting to develop epidemiological models that simultaneously track multiple diseases with similar or confounding symptoms, such as CoViD-19 and ILI (Influenza-like illnesses) as it would allow policy models e.g. testing and quarantining models to be checked for false positive and missed detection rates. Lastly, non-linear systems with negative feedback loops such as epidemiological models, are known to exhibit chaotic behavior \citep{Bolker1993,EilersenJS2020}. It is interesting how techniques like \myalgo can be adapted to handle such systems.

\subsection{Relevance to CoViD-19 and Future Prospects}
\label{sec:disc}
Given the evolving nature of the current CoViD-19 pandemic, a technique like \myalgo helps in optimally designing multi-phase lock-downs, thus avoiding speculation and human error. As shown in Sec~\ref{sec:exps}, whenever multiple waves of infection are unavoidable due to constraints on lock-downs, \myalgo offers lock-down schedules that balance the peaks of these multiple outbreaks, ensuring no peak is too high. To maximize the impact of methods such as \myalgo, close interaction and collaboration is needed with experts in the epidemiological and social sciences to better align \myalgo with professional epidemiological forecasting and economic forecasting models. \myalgo's interaction with these models is of a \emph{black-box} nature which makes integration smoother and simpler.

\section*{Acknowledgements}
D.D. is supported by the Visvesvaraya PhD Scheme for Electronics \& IT (FELLOW/2016-17/MLA/194). P.K. thanks Microsoft Research India and Tower Research for research grants.

\section*{Conflict of Interest}
The authors declare that they have no conflict of interest.

\section*{Code Availability}
All code used for this study is available at the following GitHub Repository\newline
\url{https://github.com/purushottamkar/esop}

\bibliographystyle{plainnat}
\bibliography{refs}

\end{document}